\documentclass[prd,aps,twocolumn,showpacs,superscriptaddress,floatfix]{revtex4-1}
\usepackage{bm}
\usepackage{amsmath}
\usepackage{txfonts}
\usepackage{graphicx}
\usepackage{dcolumn}
\usepackage{bm}
\usepackage{color}
\usepackage{multirow}
\usepackage{subcaption}
\def\comment#1{}

\def\slashchar#1{\setbox0=\hbox{$#1$}           
   \dimen0=\wd0                                 
   \setbox1=\hbox{/} \dimen1=\wd1               
   \ifdim\dimen0>\dimen1                        
      \rlap{\hbox to \dimen0{\hfil/\hfil}}      
      #1                                        
   \else                                        
      \rlap{\hbox to \dimen1{\hfil$#1$\hfil}}   
      /                                         
   \fi}                                         %

\begin{document}
\title{Critical assessment of the equilibrium melting-based, energy distribution theory of supercooled liquids and application to jammed systems}

\author{Nicholas B. Weingartner$^{*}$}
\affiliation{Institute of Material Science and Engineering, Washington University, St. Louis, MO 63130, U.S.A.}
\affiliation{Department of Physics, Washington University, St. Louis,
	MO 63130, U.S.A.}
\email{weingartner.n.b@wustl.edu}
\author{Chris Pueblo}
\affiliation{Institute of Material Science and Engineering, Washington University, St. Louis, MO 63130, U.S.A.}
\affiliation{Department of Physics, Washington University, St. Louis,
	MO 63130, U.S.A.}
\author{K. F. Kelton}
\affiliation{Institute of Material Science and Engineering, Washington University, St. Louis, MO 63130, U.S.A.}
\affiliation{Department of Physics, Washington University, St. Louis,
	MO 63130, U.S.A.}
\author{Zohar Nussinov$^{\dagger}$}
\affiliation{Institute of Material Science and Engineering, Washington University, St. Louis, MO 63130, U.S.A.}
\affiliation{Department of Physics, Washington University, St. Louis,
	MO 63130, U.S.A.}
\affiliation{Department of Condensed Matter Physics, Weizmann Institute of Science, Rehovot 76100, Israel}
\email{zohar@wuphys.wustl.edu}

\date{\today}

\begin{abstract}
Despite decades of intense study, the mechanisms underlying the extraordinary dynamics of supercooled liquids as they approach the glass transition remain, at best, mis-characterized, and at worst, misunderstood.  A long standing endeavor is to understand the remarkable increase of the viscosity with supercooling. Recently, a new theory of supercooled liquids has been proposed that starts from first principles, using elementary statistical mechanics arguments, to derive a form for the viscosity that contains only a single fitting parameter in its simplest form. In this we demonstrate that this exact same form may be derived from a different starting point, and then critically examine its performance. In the process we find that functional form proposed fits the viscosity data of a diverse group of 45 liquids exceptionally well over a wide temperature range, and uncover a number of interesting correlations of the single parameter with various thermodynamic quantities, ultimately allowing for the prediction of low temperature viscosity from high temperature data. Additionally, we find that similar physical reasoning can be used to derive a similar, single parameter form for the viscosity of hard-sphere/jammed liquids. We demonstrate that this form accurately reproduces the viscosity of hard-spheres, suggesting an underlying universality in metastable dynamics. \end{abstract}


\maketitle

\section{Introduction.}
The glass transition remains one of the most intensely studied and debated phenomena in physics, chemistry, and materials science \cite{1,2,3,4,5,6}. Uncovering the underlying mechanism of glassy behavior would represent not only a fundamental advance in modern physics, but also would facilitate the ability to better exploit the glass-formation process. This would inevitably lead to the more efficient processing of existing glasses, as well as the production of new types of glass with novel applications. By comparison to their crystalline counterparts, glasses enjoy substantial advantages, e.g., \cite{application1,application2,application3}. These have led to numerous applications in fields as diverse as pharmaceuticals, semiconductors, biomaterials, optical recording, and many others  \cite{application3, application4,application5,application6}. 

A material in its liquid phase is distinguished from a solid by an irregular, non-ordered molecular arrangement, and an associated ability to make large-scale molecular rearrangements in response to fluctuations and perturbations. These rearrangements, known as flow, allow a liquid to relax imposed stresses and deform inelastically. This fundamental property of liquids is quantified by the viscosity ($\eta$), a dynamical measure of a liquid's resistance to flow. The viscosity of a liquid measures the ``stickiness" of the local molecular interactions, and as such, is a temperature ($T$) 
dependent variable.

Liquids in thermal equilibrium at temperatures above their melting point, $T_{melt}$ (or, more precisely, the ``liquidus temperature'' $T_l$), have temperature-dependent viscosities which are well described by 
$\eta_{equilibrium} = \eta_0 e^\frac{G(T)}{k_B T}$.
Here, $\eta_0$, the extrapolated infinite temperature viscosity, and $G(T)$, a Gibbs free energy barrier, are material dependent parameters. The interpretation of the Arrhenius form is that there exists a barrier to molecular rearrangement, which must be overcome by a thermal fluctuation of the appropriate size in order for the rearrangement to proceed. The Gibbs free energy barrier $G(T)$ is, in general, very weakly dependent on temperature above the liquidus, and is typically taken to be a constant, $E$, such that the viscosity is quite accurately described by a standard Arrhenius form
\begin{eqnarray}
{\eta_{equilibrium}=\eta_0 e^\frac{E}{k_B T}}.
\label{arrhenius}
\end{eqnarray}
As the temperature of the liquid is lowered (cooled) quasistatically, such that equilibrium is approximately maintained, the viscosity maintains the form of Eq.(\ref{arrhenius}), exponentially increasing with decreasing temperature. When the temperature reaches $T_{melt}$, the liquid reaches the limit of stability, its free energy crossing that of a solid phase \cite{explain-mix}. At this temperature, the liquid typically gives off a characteristic latent heat. In pure systems, the liquid transforms into a crystal, acquiring long-range structural order and losing its ability to flow. By contrast, when properly computed
in the idealized limit of vanishing shear, the viscosity of a solid is infinite \cite{ddd}. As alluded to above, in what follows we will take `melting' to correspond to the liquidus, the temperature at which nucleation first becomes thermodynamically favorable, and the character of the system begins to change. In actuality, we refer to the \textit{temperature at which the dominant crystalline phase} begins to nucleate, ignoring certain pathological systems where small molar additions of a minor phase can greatly alter the liquidus with little to no change in the viscosity of the liquid.

Because nucleation and growth of the crystalline phase are kinetically controlled, there is an intrinsic time dependence to crystallization. Therefore, a liquid that is cooled sufficiently quickly through $T_{l}$, may bypass crystallization and be `supercooled' to a state of metastable equilibrium at temperatures beneath the melting point. As the temperature of the supercooled liquid drops further, its viscosity begins to increase dramatically, by as much as 16 decades over a temperature interval as small as a few hundred Kelvin (this is very clearly demonstrated by the behavior o-terphenyl \cite{OTP} and many other ``fragile''  \cite{Angell,Angell1} glass formers). Eventually, a ``glass transition'' temperature ($T_g$) is reached where the viscosity is so large that molecular rearrangements cease on physically meaningful timescales, and the supercooled liquid is termed a glass. Despite the appearance of various thermodynamic signatures \cite{Cavagna} the glass transition appears to be kinetic in nature; the glass transition marks the point at which the timescale of atomic rearrangement (relaxation time) exceeds the relevant experimental timescale and the liquid falls out of equilibrium. In crystalline solids, the dynamical arrest (infinite viscosity) is due to long range structural order that appears with a first order phase transition at melting. Glasses, however, lack the long-range order customarily associated with the stiffness/rigidity of solids, instead possessing amorphous particle arrangement. Understanding the glass transition, then, requires first understanding the temperature dependence of the viscosity of supercooled liquids. It is important to stress that notwithstanding their amorphous character, the formation of structural glasses does not rely on externally imposed disorder. All conventional liquids (and possibly even superfluid Helium and other quantum fluids \cite{He1,He2,He3,He4,He5}) may be quenched into amorphous glassy structures by rapid supercooling. 

If the rapid rise of the viscosity below the liquidus temperature $T_{l}$ were simply described by the same Arrhenius form as in Eq. (\ref{arrhenius}), or even an Arrhenius form with different constant energy barrier, $E'$, then the glass transition would not be so mysterious. Decreasing the temperature would simply remove more and more kinetic energy from the molecules leading to an ever increasing scarcity in barrier-crossing events that enable molecular rearrangements. The reduction in molecular motion, then, gives the appearance of rigidity on realizable timescales, but is an entirely kinetic phenomenon. This simple behavior is not the case, however, as all glass formers depart from the well-understood form of Eq. (\ref{arrhenius}).

Complicating matters further is the fact that different glass forming liquids display a wide spectrum of `super-Arrhenius' behaviors. These are reflected by an increase of the viscosity (as $T$ is lowered) that may be far more dramatic than that predicted by Eq. (\ref{arrhenius}). Some supercooled liquids are approximately Arrhenius, whereas others show drastic departure from the Arrhenius form. There exists a broad array of liquids having behaviors in between these extremes. Long ago, Angell defined a parameter, called  ``fragility'' \cite{Angell,Angell1}, that quantifies the degree of departure from Arrhenius behavior, as well as a classification scheme for the spectrum of behaviors. Arrhenius liquids are called strong (i.e., possess a low fragility value) whereas liquids with large departures are termed fragile (high fragility value). It is widely accepted that fragility is a significant parameter characterizing the glass transition, and it is believed that fragility may correlate with both structural and dynamic phenomenon \cite{Angell,Angell1,Martinez,Bendert,structuref,gilles1}. Therefore, any reasonable theory of glass formation and supercooling, must, at the very least, contain a connection with fragility.

The much celebrated Vogel-Fulcher-Tamman (VFT) form \cite{Vogel,Fulcher,Tamann},
\begin{eqnarray}
\label{VFT}
\eta = \eta_0 e^\frac{DT_0}{T-T_0} ,
\end{eqnarray}
has been shown to provide a more reasonable fit than the Arrhenius form to the viscosity of most supercooled liquids over a moderate range of temperatures, and has been in use since the 1920s \cite{Vogel,Fulcher,Tamann}. As seen in Eq. (\ref{VFT}), the VFT form contains three material-dependent fitting parameters, the prefactor $\eta_0$, the constant $D$, and the temperature $T_0$. These parameters are generally not predictable from first principles. Additionally, despite its apparent successes, the VFT form suffers from two fundamental drawbacks. First, it is a purely empirical function, it is not derived from first principles, or any specific theories of glass formation (although it can be reproduced by certain theories, see \cite{Biroli,Angell,Procaccia,Kalogeras,Hunter,Langer,AdamGibbs}). Secondly, it predicts a dynamic singularity at the temperature $T_0$ which exists beneath the glass transition. This temperature has been shown to be in rough agreement with the Kauzmann temperature associated with hypothesized vanishing of configuration entropy \cite{Kauzmann}, leading some to postulate that there exists a true equilibrium thermodynamic phase transition in the limit of infinitely slow cooling to $T_0$. While the notion that the slow dynamics near $T_g$ is associated with the ``ghost" of an underlying phase transition is compelling, any experimental evidence suggesting an ``ideal glass transition" remains hidden \cite{Darkly}. There are, in fact, several experimental indications that the VFT and similar forms are incorrect, e.g., \cite{amber}.

As the above discussion hints, rationalizing the mysterious super-Arrhenius increase of the shear viscosity of supercooled liquids has long been an open fundamental problem \cite{paw}. Towards that end, numerous theories have been proposed that aim to reproduce the behavior of the viscosity upon cooling, and to provide a physical framework that explains the rich phenomenology associated with the glass transition \cite{Biroli,Angell,Procaccia,Kalogeras,Hunter,Langer,Kauzmann,Cavagna}. Many such theories have been proposed and tested, all to varying degrees of success \cite{Biroli,Angell,Procaccia,Kalogeras,Hunter,Langer,AdamGibbs,Cavagna,Kirkpatrick1,Kirkpatrick2,Kirkpatrick3,Kirkpatrick4,Kirkpatrick5,Kirkpatrick6,Reichman+,Ritort}. In many cases, the functional forms derived for the viscosity are not universal; models that accurately describe metallic liquids may not work well for silicate liquids, for example \cite{Angell3}. A complete theory of the supercooled liquids, then, should answer the two most fundamental questions in the field, namely, (i) what is the cause of the super-Arrhenius viscosity and what functional form is most accurate, and (ii) is this form universal to all supercooled liquids regardless of fragility, bonding type, etc. 

Recently, a new framework for understanding the  behavior of supercooled liquids was introduced based on the framework of equilibrium statistical mechanics \cite{Nussinov,ESDH}. As we will briefly
review next in Section \ref{sec:NEDH}, this approach relies on the characteristics of the quantum eigenstates \cite{Nussinov} or, correspondingly, on the features of the classical microstates \cite{ESDH} of fixed energy to describe the phenomenology of supercooled liquids and glass formation. The resulting prediction for the viscosity was briefly demonstrated to be quite accurate for liquids of all types and fragilities at all temperatures below their respective liquidus temperatures \cite{ESDH}; we succinctly reported on the experimental collapse (that is implied by this prediction) of all available viscosity data onto a universal curve. In the current work, we will critically examine the statistical performance of the predicted form for the viscosity in this approach and extend it to temperatures above the liquidus. Additionally, we will perform a detailed analysis of the physical meaning of the single parameter, and demonstrate that it correlates strongly with various thermodynamic and dynamic quantities. Using these relations, we will demonstrate that it is possible to predict the viscosity at low temperatures, based solely on high temperature data. We will conclude by suggesting extensions of this framework to non-thermal transitions, such as jamming, and show that the framework can quite accurately capture athermal ``glassy" dynamics. 

\begin{figure}
\includegraphics[width=1 \columnwidth, height=.35 \textheight]{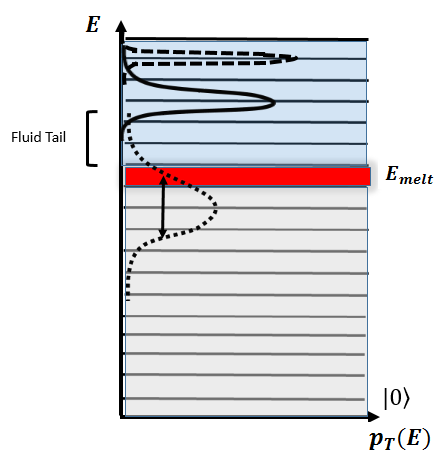}
\caption{(Color Online.) Simple pictorial representation of the basic DEH principle. This simple picture represents the spectral `hierarchy' of energy eigenvalues, $E$, and their associated eigenstates (horizontal lines), with the various probability densities, ${\it{p_{T}}}(E)$, overlaid.  Distributions shown for (i.) the high temperature equilibrium liquid (narrow dashed curve), (ii.) a high temperature non-equilibrium liquid (solid curve), and (iii.) a supercooled liquid state (dotted curve). For the dotted curve, note that the tail in the fluid-like states is what determines the hydrodynamic relaxation rate (viscosity).}
\label{eigen.}
\end{figure}

\section{The Non-equilibrium Distribution Hypothesis}
\label{sec:NEDH}

The crux of our approach is that {\it the very same quantum eigenstates \cite{Nussinov} or classical microstates \cite{ESDH}} that appear in equilibrium averages suffice
to describe supercooled liquids and glasses. That is, to describe glasses, perhaps one need not think about novel states or exotic transitions of one special sort
or another but rather employ standard statistical mechanics. The initial impetus to consider such a (seemingly all too simple) possibility was triggered by a general argument.
Specifically, in the quantum arena, the eigenstates of the disorder free Hamiltonian (describing the equilibrium solid and liquid) form
a complete basis. Thus, any state or probability density (whether that of the equilibrium system or
of the non equilibrated supercooled liquid) describing the system may be expanded in terms of
this complete set of eigenstates \cite{Nussinov}. Similar considerations may be enacted,
{\it mutatis muntandis}, for classical phase space states \cite{ESDH}. Thus, in our minimal framework,
{\it the only difference} between (i) equilibrium liquids and 
solids and (ii) the non-equilibrated supercooled liquids and glasses is that of {\it the probability distributions} that sample this complete set of states 
are different in both cases; the central idea is that the probability distributions (and long time averages) associated
with supercooled liquids and glasses may be rather trivially expressed in terms of those of the equilibrated system. 
This hypothesis of a trivially extended probability distribution in the energy density (that includes the equilibrated system as a special instance of this distribution that is of vanishing width) underlies our work. In what follows, we summarize several key aspects of the original (quantum) approach of \cite{Nussinov} which we dub the ``Distributed Eigenstate Hypothesis'' (DEH). A direct classical dual of this description (that of the ``Energy Shell Distribution Hypothesis'' (ESDH)) was introduced in \cite{ESDH}. Thus, notwithstanding the viable importance of quantum effects
\cite{qe,NussTome} in 
non cryogenic fluids, we wish to underscore that the
results that we will empirically test in great detail in the current work {\it do not}, at all, hinge on quantum mechanics. 
Planck's constant does not appear in our results. The upshot of the below review of \cite{Nussinov} is the prediction of Eq. (\ref{Final}). 
This prediction for the viscosity and the universal viscosity collapse that it implies will be assessed and extended in the next Sections. 

\subsection{Averages within the DEH theory} 
\label{deh1}
The evolution of a general $N$-body system is governed by its Hamiltonian, ${\mathcal{H}}$. For `ordinary' glass forming liquids (i.e., systems not requiring ad hoc, `quenched disorder' \cite{Cavagna}) behaving non-relativistically, the correct many body Hamiltonian consists of the kinetic energies of all $N$ particles, as well as the complete set of electrostatic interactions between all nuclei and electrons in a volume $V$. Quite generally, then, one can write down the \textbf {\textit{exact}} Hamiltonian of such a general $N$ body system \cite{Nussinov}. As always, the dynamic evolution of the system, then, is entirely determined by solving the Schr\"odinger equation corresponding to this Hamiltonian, 
\begin{eqnarray}
\label{Eigen}
\mathcal{H} |\phi_n \rangle=E_n  |\phi_n \rangle,
\end{eqnarray}
and evolving according to the appropriate time evolution operator. For macroscopic systems, $\mathcal{H}$ contains an astronomical number of interaction and kinetic terms. While methods for obtaining approximate solutions exist, in general, the exact eigenstates and associated energies cannot be determined. In spite of this apparent complication, the mere existence of the eigenstates of Eq. (\ref{Eigen}) (guaranteed by the postulates of quantum theory) allows us to make powerful statements about the dynamical and thermodynamic properties of the system. In order to make these statements, we need only rely on simple, well documented, physical characteristics of the macroscopic thermodynamic properties of equilibrium materials. We will employ said observations at various temperatures and utilize basic statistical mechanical principles.

In the following calculations, we will assume that our many body system is approximately isolated, and therefore employ the microcanonical (m.c.) ensemble. Computing measurable thermodynamic values for equilibrium systems in the microcanonical ensemble involves taking phase space averages over states within a narrow range of energies (an effective ``energy shell") that are consistent with the external constraints. When considering quantum mechanical systems, the energy levels are quantized, and the energy shell encompasses some subset of the allowed eigenstates of the system. 
Within the microcanonical ensemble, the equilibrium average of any observable ${\mathcal{O}}$ is given by 
\begin{eqnarray}
\label{dictionary}
\langle {\mathcal{O}} \rangle_{m.c.} = \frac{1}{{\cal{N}}[E-\Delta E,E]} \sum_{E - \Delta E \le E_{n} \le E} \langle \phi_n| {\mathcal{O}} | \phi_n \rangle,
\end{eqnarray}
where ${\cal{N}}[E-\Delta E,E]$ is the number of eigenstates having energies $E - \Delta E \le E_{n} \le E$ with $\Delta$ an arbitrary (system size independent) width. 
In the limit in which the width $\Delta E$ of the energy shell is made vanishingly small, only a single eigenstate (or a set of degenerate eigenstates) is effectively encompassed. If only a single, or small range of eigenstates are being averaged over, then the observed thermodynamic values must be properties of the eigenstate(s). Essentially, all observed/measured thermodynamic properties of the macroscopic, equilibrium system will then correspond to such eigenstates. This implies, as we next elaborate on, that we can use experimental observation (that provides the lefthand side of Eq. (\ref{dictionary})) to classify the properties of allowed eigenstates at various energies (as implied by the average on the righthand side of Eq. (\ref{dictionary})). It has been empirically known for millennia, that at sufficiently high energies, the equilibrium state of most materials is a liquid (with all the properties therein), whereas the low energy equilibrium state is that of a solid (with associated properties). It can, therefore, be reasonably hypothesized that, on average (in the sense implied by the righthand side of Eq. (\ref{dictionary})), the many-body eigenstates will, respectively, exhibit `liquid-like' or `solid-like' characteristics, for states with respectively high or low energy densities. Further, as it is observed that liquids and crystalline solids are separated by a first order melting/freezing phase transition, the liquid-like states are separated from the solid-like states by a 'band' of eigenstates corresponding to this melting phase transition range. The width of this melting band corresponds to the latent heat of the associated phase transition. We refer to the states in the melting band as the `Phase Transition Energy Interval' (${\cal{PT}}E{\cal{I}}$) \cite{Nussinov}. These states display properties associated with both liquid and solid states. The eigenstates associated with liquids and solids will necessarily possess the observed equilibrium structures of the respective states, in addition to the thermodynamic and kinetic properties. As such, the liquid-like eigenstates will be delocalized in the sense that a system existing in one of these eigenstates can ergodically explore phase space, and will be capable of hydrodynamic flow. Conversely, the solid-like eigenstates will be localized, breaking ergodicity, and possessing the rigidity of the crystalline solid and lacking the ability to flow. Physically, this means that each eigenstate will possess a characteristic structural relaxation time and associated visco-elastic properties, with the solid-like states being assumed ideal (infinite relaxation time). Additionally, the symmetry breaking solid-like states will necessarily have the long-range structural ordering of the equilibrium crystal built in, and the spectrum of `excited states' lying between the ground state (absolute zero) and the melting band will correspond to various phonon modes.

As discussed above, an isolated equilibrium system explores states within an infinitesimally narrow band of energy densities, and these states possess the system properties. Therefore, we can approximate the equilibrium `distribution' of energies as a delta function, $\delta(E'-E)$ peaked at the external energy. Starting at high energy, or temperature, (energy and temperature are simply related via the heat capacity) and \textbf{quasistatically} lowering the energy/temperature (cooling) will cause the system to transition to eigenstates of progressively lower energy, with the distribution remaining effectively $\delta$-peaked at the appropriate lower energies. At the ${\cal{PT}}E{\cal{I}}$, the system transitions through the mixed states, giving off latent heat, and eventually undergoes the usual first order transition. Consequently, the system then moves into a crystalline, solid-like eigenstate, possessing all of the thermo-mechanical properties of a crystalline solid. As confirmed by experiment, this only happens when equilibrium is maintained. If, instead, the system is rapidly quenched by strongly coupling to a heat bath of some kind, the system will be driven from equilibrium. The quench can be represented by a perturbing Hamiltonian, $\mathcal{H}'(t)$. The augmented Hamiltonian, $\mathcal{H}_{Full} = \mathcal{H} + \mathcal{H}'(t)$, will {\textit{not}} commute with $\mathcal{H}$ as the system energy (since the expectation value of $\mathcal{H}$ is lowered); the supercooled system will be driven into a new state, $|\Psi_T \rangle$ (where $T$ corresponds to the temperature the system was quenched to). This new state will not be an equilibrium eigenstate of the original Hamiltonian, but because the eigenstates of the original Hamiltonian, are complete, the new state can be expanded in terms of them. Therefore, the post-quench supercooled state, $|\Psi_T\rangle$, can be generically expanded in the basis of equilibrium eigenstates of the original Hamiltonian, $\mathcal{H}$, taking the form  
\begin{eqnarray}
\label{Super}
|\Psi_T \rangle = \sum_n c_n |\phi_n \rangle.
\end{eqnarray}
We see, then, that the effect of the perturbation (quench) is to \textit{mix} the eigenstates of varying energy densities $E_n/V$, such that the system to no longer exists in a single equilibrium eigenstate. The mixed state encompasses a `metastable' \textbf{distribution} (no longer a delta function) of multiple equilibrium eigenstates; this is a defining property of the Distributed Eigenstate Hypothesis \cite{Nussinov}. Realistically, there exists a density matrix associated with an open quantum system. To provide the simplest quintessential account of the theory, in what follows, we do not elaborate on the complete treatment involving the full density matrix, essentially focusing on typical states $| \Psi_T \rangle$ that are of high probability (i.e., we will consider those eigenstates of the density matrix for which the corresponding eigenvalues are high). Consideration of the full density matrix will not impact the final results reviewed herein \cite{Nussinov}. In the simple single state account, the squared amplitudes $\{{|c_n|}^2\}$ represent the probability distribution of the eigenstates of Eq. (\ref{Eigen}). Assuming that the many-body eigenstates are vanishingly close together in their energy densities, we will take the continuum limit of this distribution, $|c_n|^2 \rightarrow {\it{p}}_{T}(E')$. A similar distribution ${\it{p}}_{T}(E')$ will appear when the full density matrix is considered. A cartoon of this continuum limit probability distribution, ${\it{p_{T}}}(E')$, for various temperatures is depicted in Figure (\ref{eigen.}). As discussed above, the equilibrium state of both the liquid and solid will correspond to a probability density which is ``$\delta$-function peaked'' (see the dashed curve in Figure (\ref{eigen.})) and has its support only over a narrow shell of eigenstates that either share the same energy or are nearly degenerate. By comparison to the equilibrated system, the supercooled system exhibits a broadened probability density (see the solid and dotted curves in Figure \ref{eigen.}) which encompasses many of the equilibrium eigenstates. Long time average (l.t.a.) values will now read \cite{Nussinov},
\begin{eqnarray}
 \mathcal{O}_{l.t.a.} =\int dE'~ {\it{p}}_{T}(E') \mathcal{O}(E').
\label{average}
\end{eqnarray} 
In general, the probability density
${\it{p}}_{T}(E')$ will have a weight originating from both the higher energy delocalized liquid-like states, and the lower energy localized solid-like states. Our central thesis is that the `mismatch' of characteristics from different types of states is what leads to the phenomenology of supercooled liquids. Generally, the probability density will shift downward as the temperature is lowered, and the width of the distribution will also vary with temperature. Only high energy ``liquid-like'' equilibrium eigenstates (i.e., states $| \phi_n \rangle$ having an energy density $E_n/V$ exceeding that of the melting energy) contribute \textit{appreciably} to the long time average associated with fluid flow. 
Setting $\mathcal{O}$ in Eq. (\ref{average}) to be the vertical velocity operator $v$ of a freely falling sphere in the supercooled liquid, recognizing that the terminal speed of free fall in an equilibrium solid is zero, and invoking the Stokes relation between the terminal speed and the viscosity ($v_{l.t.a.} \propto 1/\eta$), we find that \cite{Nussinov}
\begin{eqnarray}
\label{viscosity}
\eta(E) \simeq \frac{\eta^{eq}(E_{melt})}{  \int_{E_{melt}}^{\infty} ~{{\it{p}}}_{T}(E')~ dE'} \quad | \quad  T \leq T_{l}.
\end{eqnarray}
Here, $\eta^{eq}(E_{melt})$ is the value of the viscosity of the equilibrium liquid infinitesimally above its melting temperature 
Since a reduction of flow sets in at the energy density associated with the equilibrium melting/liquidus temperature, in Eq. (\ref{viscosity}), we took this point to define a lower energy cut-off. In reality, the liquidus does not mark a hard cut-off, as states in the ${\cal{PT}}E{\cal{I}}$ may enable some level of long time mobility. 

\subsection{The scale-free Gaussian distribution and the viscosity function that it implies}
\label{deh2}
\begin{figure*}
\centering
\includegraphics[width=1.8 \columnwidth]{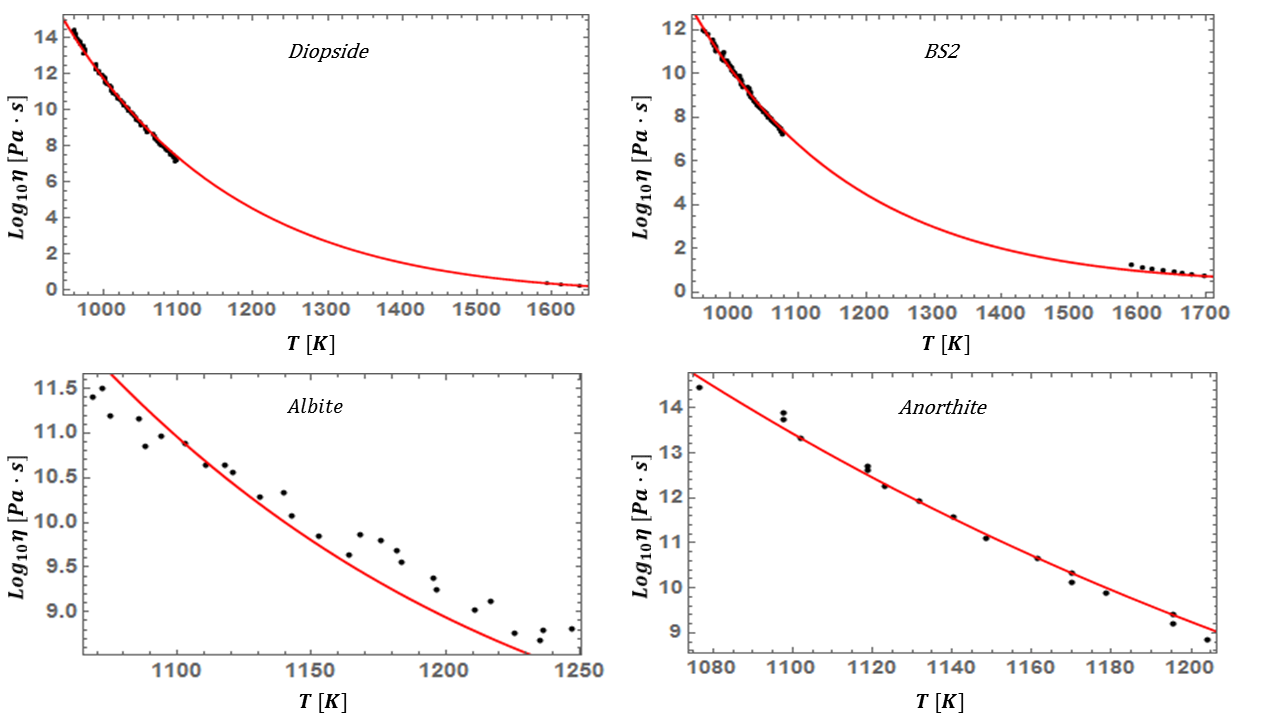}
\caption{(Color Online.) Fits to the viscosity of four sillicate glassformers with the DEH form of Eq. (\ref{Final}).}
\label{COMP.}
\end{figure*}
\begin{figure*}
\centering
\includegraphics[width=1.8 \columnwidth]{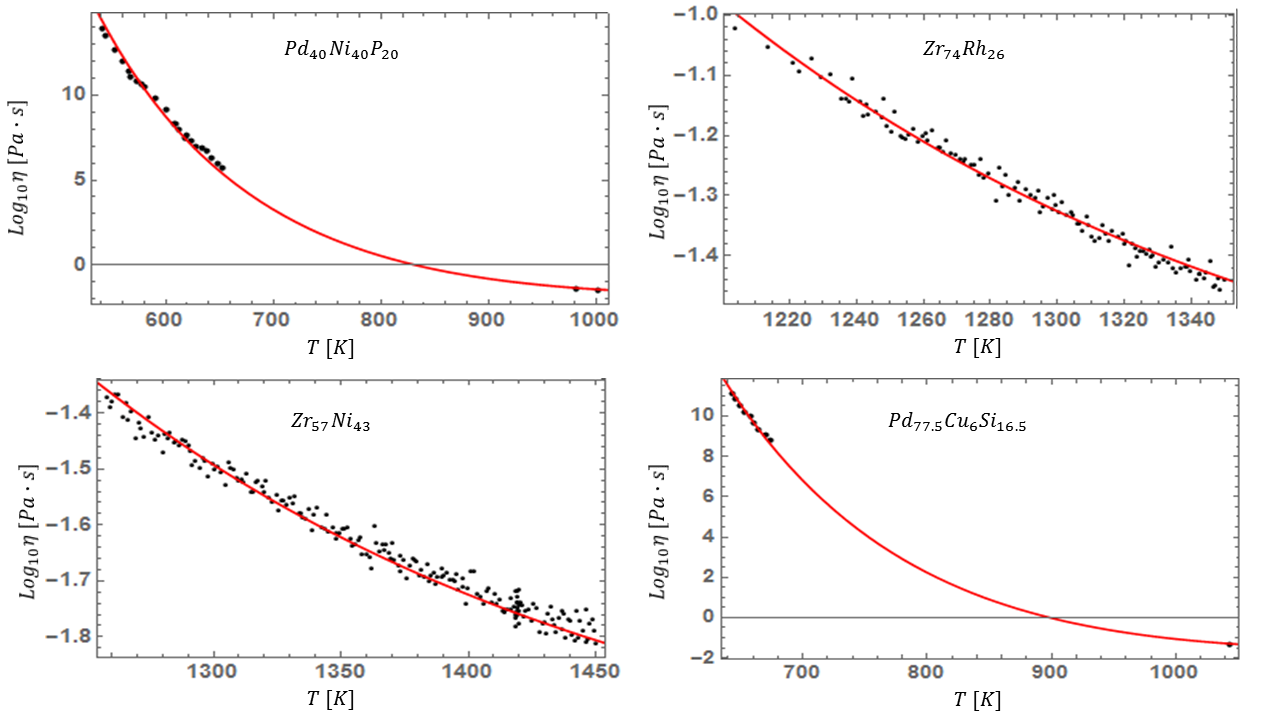}
\caption{(Color Online.) Fits of the DEH form for the viscosity, Eq. (\ref{Final}), to four metallic glassforming liquids.}
\label{Metallics.}
\end{figure*}
In order to obtain an approximate form for the viscosity (and other observables) in the DEH model, we need to have an explicit form for the distribution function, ${\it{p_{T}}}(E)$. We do not know, \textit{a priori} what this function is, but we do know characteristics it must possess: {\bf{(i)}} The distribution must be normalized. {\bf{(ii)}} The probability density must be such that the average of $ \mathcal{H}$ is equal to the measured system energy $\langle E \rangle$. {\bf{(iii)}} Since supercooled liquids are not in full thermodynamic equilibrium, the distribution cannot be a delta function in the energy density.  Thus
the distribution of the energy density must display a non-vanishing standard deviation in the thermodynamic limit. {\bf{(iv)}} In the absence of any additional information, the unknown distribution must maximize the Shannon entropy, $H_I= \int dE'~{\it{p}}_{T}(E') \log_{2}[{\it{p}}_{T}(E')]$, subject to constraints {\bf{(i)}}- {\bf{(iii)}}; the distribution maximizing the Shannon entropy subject to the constraint of a fixed standard deviation is, trivially, a Gaussian distribution. 
Thus, putting all of the pieces together, in the absence of additional information, the distribution 
\begin{eqnarray}
{{\it{p}}_{T}(E')=\frac{1}{\sqrt{2\pi \sigma_T^2}} e^{-\frac{(E'-\langle E \rangle)^2}{2 \bar{\sigma}_T^2}}}
\label{Gaussian}
\end{eqnarray} 
uniquely satisfies all of the above requisite characteristics. 
As is well known, the distribution of the energy density in the equilibrated system is a Gaussian in which the standard deviation scales as $N^{-1/2}$ (and thus vanishing in the thermodynamic
limit). Thus, Eq. (\ref{Gaussian}) is a trivial extension of the distribution present in equilibrated systems. For completeness, we remark that other Gaussian distributions have, of course, been observed before in disparate contexts. For instance, in \cite{Gaussian2} and many other works the distributions associated with
local low energy ``inherent states'' in supercooled liquids were analyzed through the prism of Gaussian distributions. To avoid confusion, we wish to stress that the probability density
of Eq. (\ref{Gaussian}) is that associated
with the energy densities of all states (not that of inherent states); we underscore that we do not consider local metastable energy minima in an ``energy landscape'' and fluctuations about them. 
\begin{figure*}
\centering
\includegraphics[width= 1.8 \columnwidth, height= 0.5 \textheight, keepaspectratio]{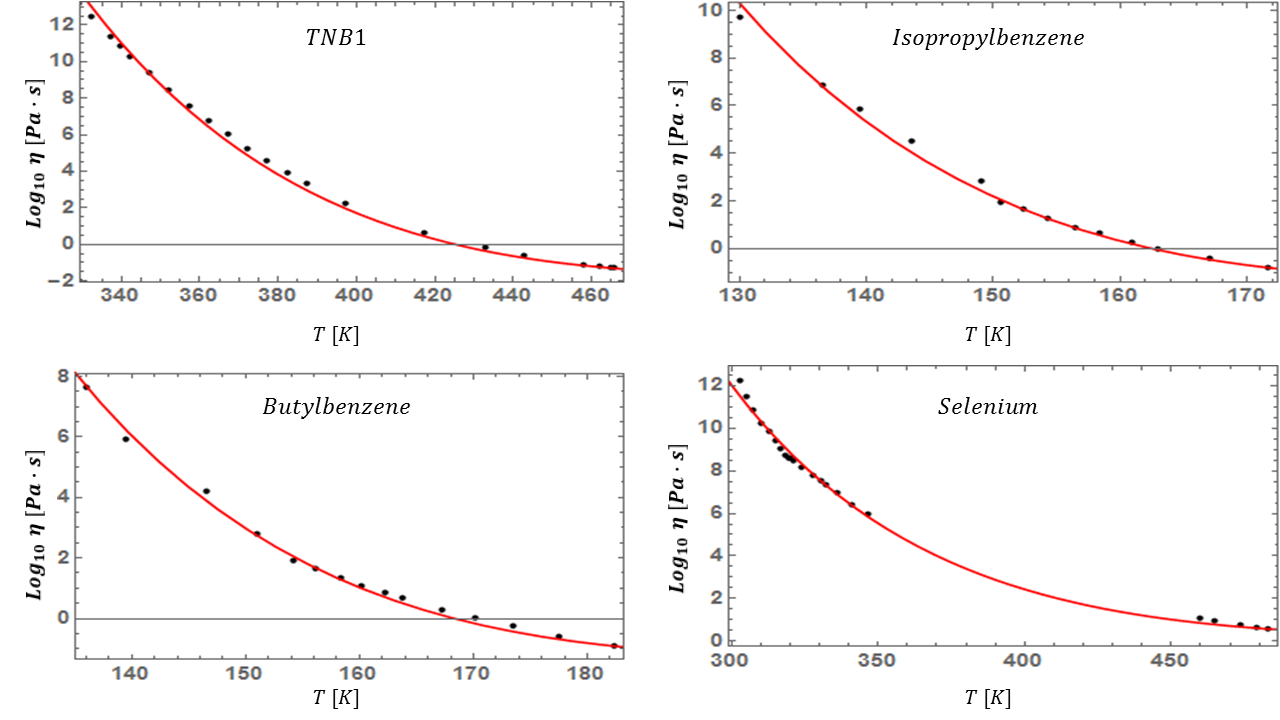}
\caption{(Color Online.) Fits of the DEH form for the viscosity, Eq. (\ref{Final}), to various benzene and chalcogenide liquids.}
\label{Chal.}
\end{figure*}
\begin{figure*}
\centering
\includegraphics[width=1.8 \columnwidth, height= .5 \textheight, keepaspectratio]{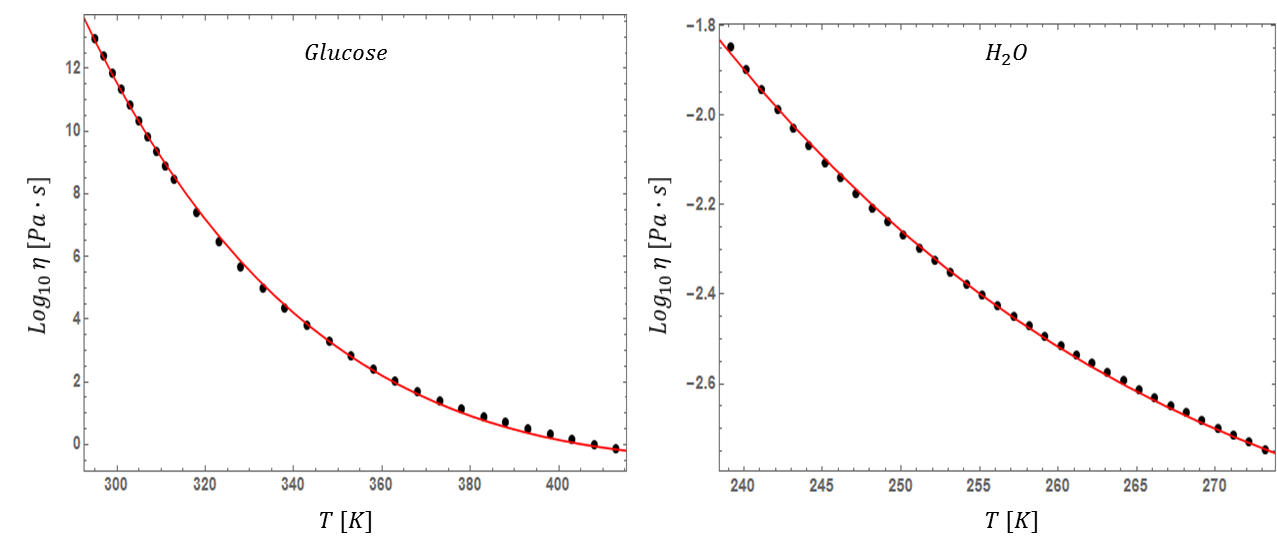}
\caption{(Color Online.) Fits of the DEH form for the viscosity, Eq. (\ref{Final}), to supercooled glucose and supercooled water.}
\label{Glucose.}
\end{figure*}

Inserting Eq.(\ref{Gaussian}) into Eq.(\ref{viscosity}) will yield the viscosity as a function of energy. In order to compare our theoretical notions to experimental data, we need to express the viscosity as a function of measured temperature (and not the energy density). Towards this end, we will define the average heat capacity in the range $[T,T_{melt}]$ given by $C(T) \equiv \frac{E_{melt}-\langle E \rangle}{T_{melt} - T}$ where $\langle E\rangle$ is the energy of the supercooled liquid at a temperature $T$. In reality, the ratio defining $C(T)$ does not change substantially as a function of temperature $T$; the function $C(T)$ does nevertheless vary with temperature yet its weak temperature will identically drop out in our final result of Eq. (\ref{Final}). We will assume that 
the dimensionless ratio
$\bar{A} \equiv \Big(\frac{\sigma_{T}(T_{melt} - T)}{T (E_{melt} - \langle E \rangle )} \Big) \equiv  \Big( \frac{\sigma_T }{ CT} \Big) \equiv \Big(\frac{\bar{\sigma}_{T}}{T} \Big)$
does not vary strongly in the interval of experimentally measured temperatures. 
An assumption of constant values of the dimensionless ratio $\bar{A}$ and $C \sim \langle E \rangle/T$ is tantamount to asserting that the Gaussian of Eq. (\ref{Gaussian}) is scale free. By this assumption of being ``scale free'', we mean only energy scale 
(whether that for the average energy $\langle E \rangle$ or the width $\bar{\sigma}_{T}$) is set by the temperature $T$. As we emphasized above, any temperature dependence of the average heat capacity defined by the ratio defining $C(T)$ will drop out in the final expression that we provide next. 
With all of the above, the viscosity of the supercooled liquid is \cite{Nussinov}
\begin{eqnarray}
\label{Final}
\eta(T) = \frac{\eta_{s.c.}(T_{melt})}{{\rm erfc} \Big( \frac{E_{melt}-\langle E \rangle}{\bar{\sigma}_{T} \sqrt{2}} \Big)} = \frac{\eta(T_{l})}{{\rm erfc}\left(  \frac{T_{l} - T}{\sqrt{2} \bar{\bar{\sigma}}_T}\right)} = \frac{\eta(T_{l})}{{\rm erfc}\left(\frac{T_{l} - T}{\sqrt{2} \bar{A} T}\right)}.
\end{eqnarray}
The first equality in Eq. (\ref{Final}) is obtained by substituting Eq. (\ref{Gaussian}) into Eq. (\ref{viscosity}). The last two equalities follow from our definitions of $C$ and $\overline{A}$.
If the ratio of $\overline{A}$ does not significantly change with $T$ in the measured temperature range then we may set it to be a constant (as we will in this work).
Alternatively, one may arrive at Eq. (\ref{Final}) by assuming that the effective temperatures $T_n$ of the equilibrated system (i.e., eigenstates having an energy $E_n= U(T_n)$ with $U$ the internal energy of the equilibrated system governed by $\mathcal{H}$) are distributed in a Gaussian fashion about the imposed external constraint that the supercooled liquid has a temperature $T$.
In the above, we largely reviewed the quantum DEH model of \cite{Nussinov} that first predicted Eq. (\ref{Final}). As we noted earlier, a derivation of the same result in the framework classical statistical physics appears in \cite{ESDH}.  

The prediction of Eq. (\ref{Final}) for the viscosity at all temperatures below the liquidus temperature, $T<T_{l}$, requires knowledge of the liquidus temperature and the viscosity $\eta(T_l)$ of the supercooled liquid at this temperature. Both of these quantities are given by experiment, and are not fitting parameters of the theory (these numerical values of these measurable quantities 
are presented in Table I). An objective of the current paper is to critically test the performance of this function that goes beyond the initial analysis conducted in \cite{ESDH}.
\begin{table*}[t]
\centering
\caption{Values of Relevant Parameters for all liquids studied}
\label{A}
\begin{tabular}{*{4}{@{\hskip 0.4in}c@{\hskip 0.4in}}} 
\toprule
\emph{Composition} \quad & \emph{}$\bar{A}$ & \emph{$T_{l}$ [K]} & \emph{$\eta(T_{l})$ [Pa*s]} \\
\colrule
BS2 & 0.111107 & 1699 & 5.570596   \\
Diopside & 0.094984 & 1664 & 1.5068     \\
LS2 & 0.12048 & 1307 & 22.198     \\
OTP & 0.049275 & 329.35 & 0.02954    \\
Salol & 0.061654 & 315 & 0.008884   \\
Anorthite & 0.092875 & 1823 & 39.81072   \\
Zr$_{57}$Ni$_{43}$ & 0.165584 & 1450 & 0.01564    \\
Pd$_{40}$Ni$_{40}$P$_{20}$ & 0.10939 & 1030 & 0.030197   \\
Zr$_{74}$Rh$_{26}$ & 0.132831 & 1350 & 0.03643    \\
Pd$_{77.5}$Cu$_6$Si$_{16.5}$ & 0.088303 & 1058 & 0.0446     \\
Albite & 0.073075 & 1393 & 24154952.8 \\
Cu$_{64}$Zr$_{36}$ & 0.101088 & 1230 & 0.021      \\
Ni$_{34}$Zr$_{66}$ & 0.148039 & 1283 & 0.0269     \\
Zr$_{50}$Cu$_{48}$Al$_{2}$ & 0.118278 & 1220 & 0.0233     \\
Ni$_{62}$Nb$_{38}$ & 0.07742 & 1483 & 0.042      \\
Vit106a & 0.094557 & 1125 & 0.131      \\
Cu$_{55}$Zr$_{45}$ & 0.102192 & 1193 & 0.0266     \\
H$_2$O & 0.094094 & 273.15 & 0.001794   \\
Glucose & 0.056183 & 419 & 0.53       \\
Glycerol & 0.076957 & 290.9 & 1.9953     \\
Ti$_{40}$Zr$_{10}$Cu$_{30}$Pd$_{20}$ & 0.13109 & 1279.226 & 0.01652    \\
Zr$_{70}$Pd$_{30}$ & 0.149006 & 1350.789 & 0.02288    \\
Zr$_{80}$Pt$_{20}$ & 0.119757 & 1363.789 & 0.04805    \\
NS2 & 0.095195 & 1147 & 992.274716 \\
Cu$_{60}$Zr$_{20}$Ti$_{20}$ & 0.073101 & 1125.409 & 0.04516    \\
Cu$_{69}$Zr$_{31}$         & 0.111355  & 1313     & 0.01155    \\
Cu$_{46}$Zr$_{54}$         & 0.110984  & 1198     & 0.02044535 \\
Ni$_{24}$Zr$_{76}$         & 0.173226  & 1233     & 0.02625234 \\
Cu$_{50}$Zr$_{42.5}$Ti$_{7.5}$  & 0.104828  & 1152     & 0.0268     \\
D Fructose       & 0.035443  & 418      & 7.31553376 \\
TNB1             & 0.053509  & 472      & 0.03999447 \\
Selenium         & 0.092503  & 494      & 2.9512     \\
CN60.40          & 0.105419  & 1170     & 186.2087   \\
CN60.20          & 0.113965  & 1450     & 12.5887052 \\
Pd$_{82}$Si$_{18}$         & 0.097314  & 1071     & 0.03615283 \\
Cu$_{50}$Zr$_{45}$A$l_{5}$     & 0.083885  & 1173     & 0.03797    \\
Ti$_{40}$Zr$_{10}$Cu$_{36}$Pd$_{14}$ & 0.097406  & 1185     & 0.0256     \\
Cu$_{50}$Zr$_{50}$       & 0.117874  & 1226     & 0.02162    \\
Isopropylbenzene & 0.052216  & 177      & 0.086      \\
ButylBenzene     & 0.060151  & 185      & 0.0992     \\
Cu$_{58}$Zr$_{42}$         & 0.093316  & 1199     & 0.02526    \\
Vit 1 & 0.07862 & 937 & 36.59823 \\
Trehalose & 0.050244 & 473 & 2.71828 \\
Sec-Butylbenzene & 0.056631 & 190.3 & 0.071 \\
SiO$_2$ & 0.06431 & 1873 & 1.196x$10^{8}$ \\
\botrule
\hline       
\end{tabular}
\end{table*}

\section{Tests of the DEH Model.}
\label{test.}
With Eq. (\ref{Final}) in hand, we next compare its predictions to measured viscosity data of various fluids (Section \ref{fv}), provide a thorough statistical study of the quality of these viscosity fits that are obtained by
this predicted form (Section \ref{sr}), compare our predicted to other prevalent fitting functions that have been used throughout the years (Section \ref{the_others}),

\subsection{Fitting of Viscosity Function}
\label{fv}
To test the validity of the DEH, we examined how well the functional form of Eq. (\ref{Final}) fit actual experimental viscosity data. We applied the DEH form to the viscosities of 45 different glass forming supercooled liquids. As the DEH is meant to be universal across all types of supercooled liquids, we selected glassformers of all classes, bonding types, fragilities, and physical and chemical features. We studied silicate liquids, organic liquids, metallic liquids, elemental liquids, sugars, chalcogenides, and even supercooled water. The experimental viscosity data for the various supercooled liquids was either measured ``in-house" by one of the authors or extracted from previously published works. The published data that was present only in graphical form was converted to tabular form using data digitization software. Nonlinear curve fitting methods and error minimization were employed to extract the best-fit value of the single parameter, $\bar{A}$. The natural logarithm of the DEH form of the viscosity (Eq. \ref{Final}) was fitted to the experimental viscosity data for all temperatures at and below the liquidus, using the experimental values for the liquidus temperature, $T_{l}$, and viscosity at the liquidus, $\eta(T_{l})$ (data presented in traditional base-10 format for consistency). In some cases, a data point was not present exactly at the liquidus temperature, requiring interpolation of the value of $\eta(T_{l})$. In most cases, the standard error in the calculation of $\bar{A}$ from the nonlinear fitting algorithm was of $\mathcal{O}(10^{-3})$. Associated p-values supported the null hypothesis that the values of $\bar{A}$ were statistically significant from zero \cite{pvalue}.

Figures (\ref{COMP.})-(\ref{Glucose.}) show the viscosity data, with DEH fit applied, in logarithmic form as a function of temperature. Qualitatively, the data in the figures suggest that the DEH form is capable of reproducing the data for the silicate, metallic, organic, and sugar-based glass forming liquids as well as supercooled water, with minimal residual error. A notable exception is the case of Albite. The closest data points to the melting point for this liquid were within approximately 100 K on both sides of the $T_{l}$. This necessitated a large interpolation of the data to extract $\eta(T_{l})$. While it was observed that the DEH fits are significantly less sensitive to changes in the value of $\eta(T_{l})$, than either $T_{l}$ itself, or $\bar{A}$, this is still likely the cause of this discrepancy. 

We have demonstrated qualitatively that the DEH form can reasonably reproduce the temperature dependence of the viscosity of a large number of supercooled liquids. However, purely visual fitting results are subjective, and the validity of a physical model must rest on objective measures. Therefore, in order to bolster the claim that the DEH model is the correct model for describing supercooled liquids, we also performed a quantitative analysis, to assess the statistical significance of our results. In what follows, we calculate various statistical measures of the goodness of fit of the DEH model to the experimental viscosity data, as well as perform a statistical comparison with previous theories. While we perform these analyses on all 45 liquids studied, we will use OTP, LS2, and $Pd_{40}Ni_{40}P_{20}$ as `case studies' throughout the next sections. These liquids are seen to be good representatives for their various classes, and will allow us to do detailed calculations with lower computational cost.

\begin{figure*}
\centering
\includegraphics[width=1.8 \columnwidth]{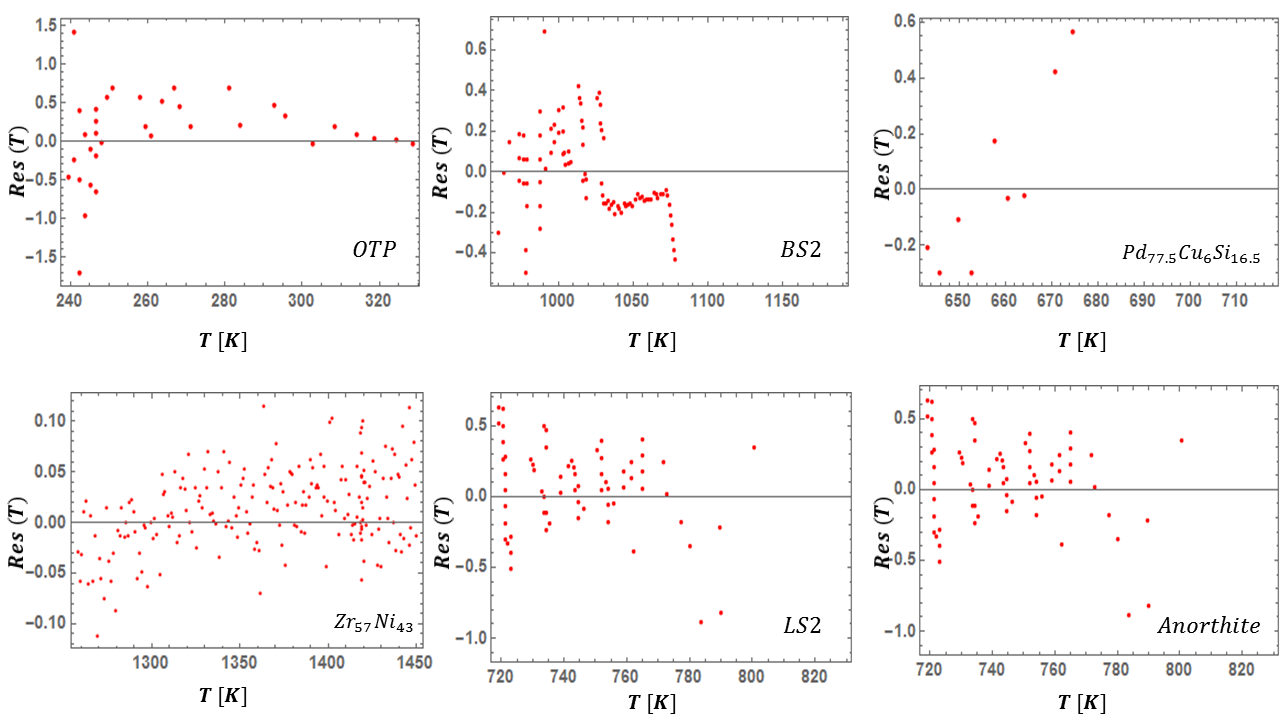}
\caption{(Color Online.) Residuals as computed from Eq. (\ref{residual}) associated with fits of Eq. (\ref{Final}) for six different supercooled liquids. Residuals corresponding to accurate fits typically possess random scatter about zero. For a discussion of possible bias in the residuals, see the main text.}
\label{Residuals.}
\end{figure*}
\begin{figure*}
\centering
\includegraphics[width=1.8 \columnwidth]{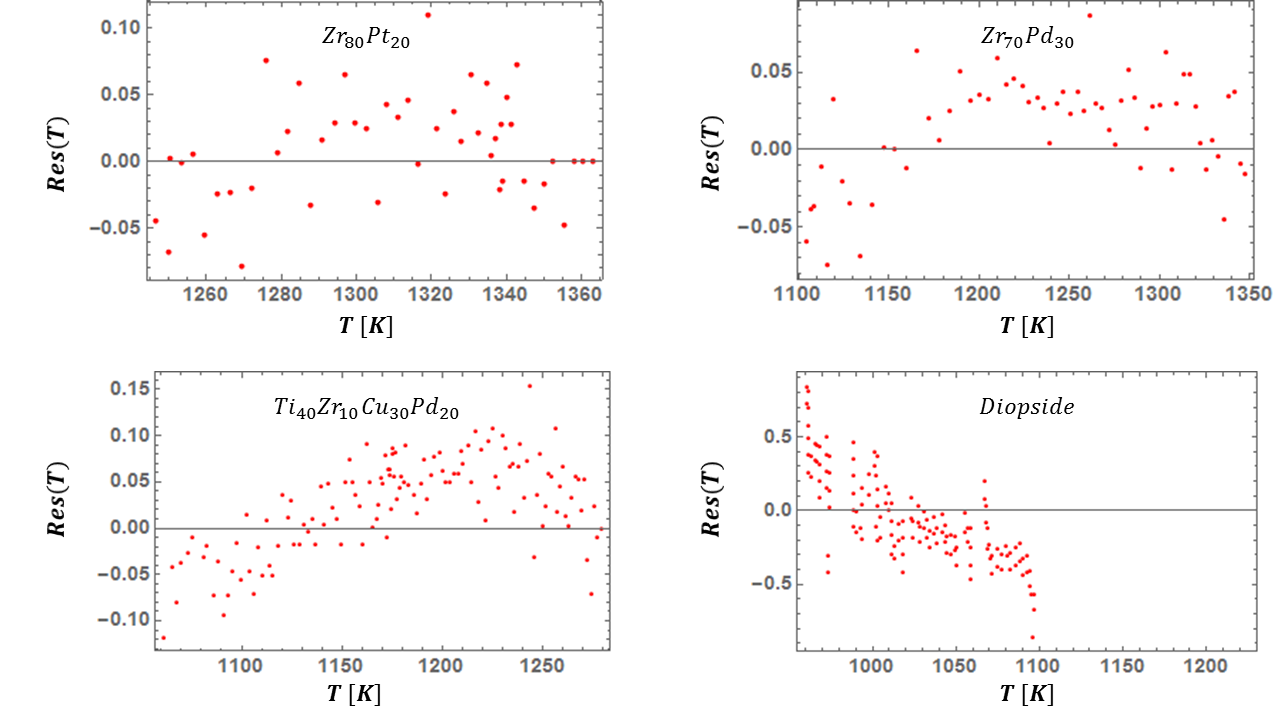}
\caption{(Color Online.)  Residuals as computed from Eq. (\ref{residual}) associated with fits of Eq. (\ref{Final}) for four different supercooled liquids which receive cross validation analysis.}
\label{Residuals2.}
\end{figure*}

\subsection{Analysis of the Statistical Residuals}
\label{sr}
Statistical measures allow one to quantify the error, or goodness of fit (GoF) of the performance of a model. Typically, the first step to assessing the statistical GoF is to analyze the residuals of the fit. In our case, using the raw viscosity data, the residuals are defined as
\begin{eqnarray}
\label{residual}
Res(T)\equiv \sum_{i} (\eta_{exp}(T_i)-f(T_i)).
\end{eqnarray}
Here, $f(T)$ is the temperature dependent model function being tested (in this section, the DEH model of Eq.(\ref{Final}). Figures (\ref{Residuals.}) and (\ref{Residuals2.}) display the results of the residual analysis for a random sample of the supercooled liquids studied. When examining the residuals of a fit, one wants to note both the magnitude of the residuals and the distribution of the residuals about zero.  In all liquids studied the magnitude of the residuals is very small, and this is true for LS2 and OTP (with the exception of two outlying points for OTP). This low magnitude suggests that the DEH form is capable of reproducing the approximate values of the measured viscosity. The distribution of the residuals about zero is a measure of how well the model captures the actual trend of the data. An accurate model should have residuals which, roughly, approximate the random error associated with measurement error. Examination of the residuals of the DEH shows that in many cases the residuals are more or less random. There are some minor exceptions, which might be due to the fact that most viscosity data is reported in the literature without error bars, and not considering these error bars in the fits is likely the culprit in the minor skewing of residuals. This hypothesis is bolstered by the fact that other models of supercooled liquids tested in the next section tend to show the same skewing, or bias in the residuals suggesting possible error inherent to the data itself.

\subsubsection{Statistical Measures of Goodness of Fit}
\label{sgf}

\begin{table*}
\centering
\caption{Statistical Measures of GoF}
\label{goodness}
\begin{tabular}{*{5}{@{\hskip 0.4in}c@{\hskip 0.4in}}}
\toprule
\emph{Composition} & \emph{Function} & \emph{$(SSE)$} & \emph{$\chi_{red}^2$} & \emph{$R^2$} \\
\colrule
\multirow{6}{2em}{OTP} & DEH & 10.617 & 0.312264 & 0.997247 \\
& VFT & 24.5584 & 0.767451 & 0.993632 \\
& KKZNT & 8.67516 & 0.279844 & 0.997751 \\
& CG & 9.66624 & 0.30207 & 0.997494 \\
& BENK & 8.91119 & 0.270036 & 0.997689 \\
& MYEGA & 12.3813 & 0.386916 & 0.99679 \\
\\
\multirow{6}{2em}{LS2} & DEH & 14.8497 & 0.215213 & 0.983678 \\
& VFT & 16.5523 & 0.247049 & 0.981807 \\
& KKZNT & 13.3202 & 0.201821 & 0.985359 \\
& CG & 14.8216 & 0.221218 & 0.983709 \\
& BENK & 13.3526 & 0.196361 & 0.985324 \\
& MYEGA & 24.9113 & 0.371811 & 0.972619 \\
\\
\multirow{6}{6em}{Pd$_{77.5}$Cu$_{6}$Si$_{16.5}$} & DEH &  0.789078 & 0.0876754 & 0.998759 \\
& VFT & 10.2834 & 1.46905 & 0.983823 \\
& KKZNT & 0.235779 & 0.0392965 & 0.999629 \\
& CG & 0.203843 & 0.0291204 & 0.999679 \\
& BENK & 0.334157 & 0.0417696 & 0.999474 \\
& MYEGA & 1.31174 & 0.187391 & 0.997937 \\
\\
{Salol} & DEH & 17.1643 & 0.553687 & 0.993136 \\
{Diopside} & DEH & 13.1776  & 0.0941259 & 0.997362 \\
{Anorthite} & DEH & 2.25807 & 0.141129 & 0.991396 \\
{BS2} & DEH & 4.902 & 0.0505361 & 0.998646 \\
{Albite} & DEH & 13.3105 & 0.511942 & 0.87503 \\
Zr$_{74}$Rh$_{26}$ & DEH & 0.115181 & 0.000984452 & 0.983959 \\
Pd$_{40}$Ni$_{40}$P$_{20}$ & DEH & 12.3782 & 0.515757 & 0.993153 \\
Zr$_{57}$Ni$_{43}$ & DEH & 0.351164 & 0.00172139 & 0.977947 \\
Cu$_{64}$Zr$_{36}$ & DEH & 0.190441 & 0.00307162 & 0.984655 \\
Ni$_{34}$Zr$_{66}$ & DEH & 0.121782 & 0.00162376 & 0.993343 \\
Zr$_{50}$Cu$_{48}$Al$_{2}$ & DEH & 10.617 & 0.312264 & 0.997247 \\
Ni$_{62}$Nb$_{38}$ & DEH & 0.448888 & 0.00487922 & 0.9841 \\
Vit106a & DEH & 6.23195 & 0.623195 & 0.996508 \\
Cu$_{55}$Zr$_{45}$ & DEH & 0.223386 & 0.00314628 & 0.987581 \\
H$_2$O & DEH & 0.00731595 & 0.000215175 & 0.999412 \\
Glucose & DEH & 1.48859 & 0.0513308 & 0.999499 \\
Glycerol & DEH & 76.0137 & 1.85399 & 0.945217 \\
Ti$_{40}$Zr$_{10}$Cu$_{30}$Pd$_{20}$ & DEH & 0.395717 & 0.00316573 & 0.988712 \\
Zr$_{70}$Pd$_{30}$ & DEH & 0.080497 & 0.00134162 & 0.996159 \\
Zr$_{80}$Pt$_{20}$ & DEH & 0.077876 & 0.00162242 & 0.971562 \\
NS2 & DEH & 20.9749 & 0.723273 & 0.981462 \\
Cu$_{60}$Zr$_{20}$Ti$_{20}$ & DEH & 0.196626 & 0.0012063 & 0.985095 \\
Cu$_{69}$Zr$_{31}$ & DEH & 0.756104 & 0.00804366 & 0.950419 \\
Cu$_{46}$Zr$_{54}$ & DEH & 0.650675 & 0.00971157 & 0.910136 \\
Ni$_{24}$Zr$_{76}$ & DEH & 0.0453595 & 0.0008584 & 0.991683 \\
Cu$_{50}$Zr$_{42.5}$Ti$_{7.5}$ & DEH & 0.0535541 & 0.00172755 & 0.982531 \\
D Fructose & DEH & 0.554086 & 0.0240907 & 0.946689 \\
TNB1 & DEH & 8.97792 & 0.448896 & 0.996155 \\
Selenium & DEH & 6.43906 & 0.292684 & 0.995906 \\
CN60.40 & DEH & 0.746426 & 0.0678569 & 0.998937 \\
CN60.20 & DEH & 0.147407 & 0.0105291 & 0.999883 \\
Pd$_{82}$Si$_{18}$ & DEH & 1.2915 & 0.1435 & 0.998916 \\
Cu$_{50}$Zr$_{45}$Al$_{5}$ & DEH & 0.109111 & 0.000742252 & 0.992842 \\
Ti$_{40}$Zr$_{10}$Cu$_{36}$Pd$_{14}$ & DEH & 0.195736 & 0.00163113 & 0.92674 \\
Cu$_{50}$Zr$_{50}$ & DEH & 0.235607 & 0.00420727 & 0.976969 \\
Isopropyl benzene & DEH & 4.47953 & 0.344579 & 0.993307 \\
Butylbenzene & DEH & 1.97384 & 0.140989 & 0.995543 \\
Cu$_{58}$Zr$_{42}$ & DEH & 0.551631 & 0.0108163 & 0.966384 \\
Vit 1 & DEH & 46.5891 & 2.58828 & 0.956556 \\
Trehalose & DEH & 8.93373 & 0.288185 & 0.934837 \\
Sec-Butylbenzene & DEH & 1.27723 & 0.159653 & 0.976809 \\
SiO$_2$ & DEH & 57.7053 & 1.98984 & 0.660326 \\
\botrule
\end{tabular}
\end{table*}
The next step in performing a rigorous statistical analysis of the quality of a model is to calculate various quantitative measures of the GoF. For each of the liquids studied we compute the sum of squared errors (SSE),
\begin {eqnarray}
\label{SSE}
SSE \equiv \sum_{i} (\eta(T_i)-f(T_i))^2,
\end{eqnarray}
r-squared value ($R^2$),
\begin{eqnarray}
\label{r}
R^2\equiv 1-\frac{SSE}{SST}=1-\frac{\sum_{i} (\eta(T_i)-f(T_i))^2}{\sum_{i} (\eta(T_i)-\bar{\eta})^2}
\end{eqnarray}
and reduced chi-squared value, $\chi_{reduced}^2$ (There are multiple definitions of the $\chi^2$ statistic for goodness of fit. We have chosen one of the more common ones),
\begin{eqnarray}
\label{chi}
\chi_{reduced}^2=\sum_{i} \frac{(\eta(T_i)-f(T_i))^2}{n_{data}-n_{parameters}}.
\end{eqnarray}
The calculated values of the statistical GoF measures are listed in Table (\ref{goodness}), in the rows labeled ``DEH". Statistically significant GoF is typically taken to correspond to $\chi_{reduced}^2$ values less than one, $R^2$ values asymptotically close to one and low values for the SSE. Examining the values in Table (\ref{goodness}) makes clear that the lowest values of $\chi_{reduced}^2$ do not always correspond to the highest values of $R^2$. In some cases the various statistical measures can report differing levels of GoF. Therefore, it is important to consider all measures simultaneously. We can examine the ``worst-case examples" to assess a bound on the DEH GoF measures. We see that the lowest value of $R^2$ corresponds to SiO$_2$ which we will discuss in more detail in the SI. Otherwise, with the exception of Albite which was discussed above, the $R^2$ values all exceed 0.9. This indicates that the DEH model is able to accurately account for the natural variability in the data.  As seen in the table, the highest values of $\chi_{reduced}^2$ are approximately 1.9 for SiO$_2$ and 1.8 for glycerol.  With these exceptions, the remainder of the liquids studied all have values of $\chi_{reduced}^2$ which are significantly less than one. This is indicative that the DEH model is capable of representing the viscosity data in a statistically significant way, while simultaneously suggesting that specific liquids may possess certain anomalies (see SI). If one were to go further and examine actual p-values associated with the various $\chi^2$ values, they would also confirm that the DEH model provides a statistically significant reproduction of the experimental data.

\subsubsection{Cross Validation}
\label{cv}
\begin{figure*}
\centering
\includegraphics[width= 1.8 \columnwidth]{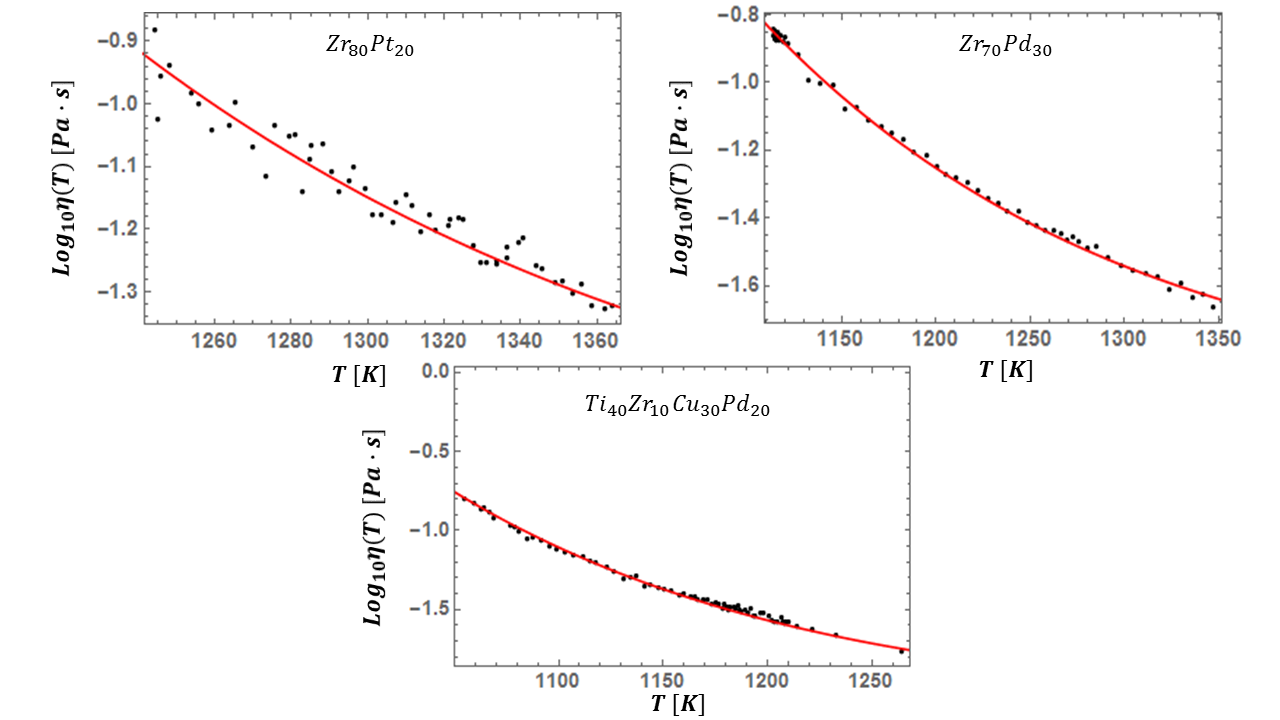}
\caption{(Color Online.)  Results of cross validation of the DEH model for three supercooled liquids where multiple data sets were available. Using values of $\bar{A}$ extracted by fitting the data of one set, we applied the form of Eq. (\ref{Final}) to a second set to assess the reproducibility of the model. For more information, see Section VC.}
\label{CV.}
\end{figure*}
\begin{figure*}
\centering
\includegraphics[width=1.8 \columnwidth]{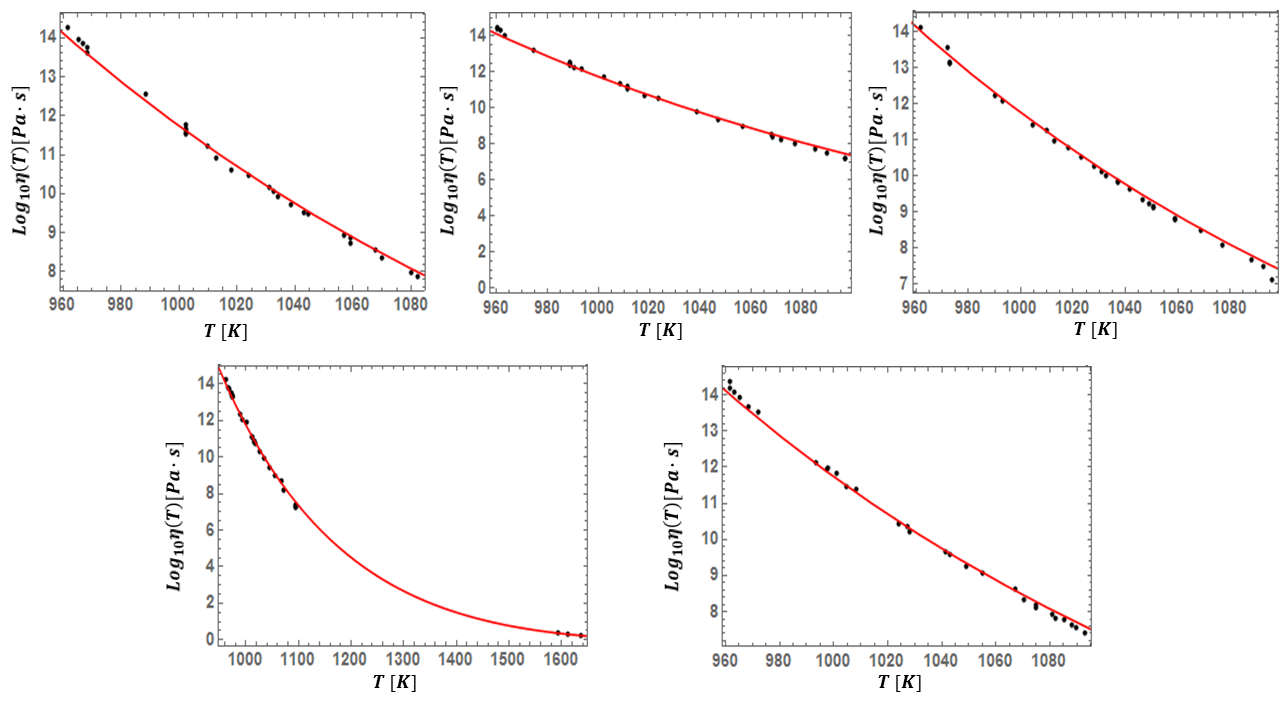}
\caption{(Color Online). Results of 5-fold cross validation of single diopside dataset. We randomly separated the single dataset for diopside into 5 subsets and iteratively used 4 subsets to fit and extract $\bar{A}$ to apply the fit of Eq.(\ref{Final}) to the fifth subset to assess the reproducibility of the model. For more information, see Section VC.}
\label{DCV.}
\end{figure*}
The statistical measures employed up to this point to assess the performance of the DEH have demonstrated that the model is capable of fitting the experimental viscosity data of a large number of supercooled liquids to a high degree of statistical accuracy. The drawback to the methods employed previously, however, is that they merely reflect the ability of the model to fit the data given, and do not describe how well the model can make predictions based on those fits or whether or not the model over fits the data. In order to assess the ability of the fits to predict `new' data for a given liquid we must employ cross validation schemes. In cross validation, we use the value of the parameter, $\bar{A}$ that comes from fitting one set of experimental viscosity data for a given liquid and apply it (using Eq. \ref{viscosity})) to an independently measured set of data for the same system. We employed this analysis for multiple independent data sets that we had available for metallic liquids. Using the values of $\bar{A}$ reported in Table (\ref{A}), we apply the DEH form to alternative data sets for three metallic liquids. The results are shown in Fig. (\ref{CV.}). For liquids that did not have multiple data sets available for cross validation we applied a 5-fold cross validation scheme to the single data set available. The example case that we examined is for diopside, and the results of this 5-fold cross validation are presented in Fig. (\ref{DCV.}). Qualitatively it is clear from the cross validation studies performed here, that the parameter values extracted from fitting experimental data sets will generalize to new data for the same liquid, providing more support that the DEH provides a statistically accurate model of supercooled liquid viscosity data. We can quantify this by examining the values of $\chi_{reduced}^2$ that result from the cross validated fits. These results are given in Table (\ref{Cross}). Comparing these values with those in Table (\ref{goodness}) for the original fits shows that they are roughly comparable in magnitude, further validating the general applicability of the DEH fit accommodate `new' data.

\begin{table}
\centering
\caption{Cross Validation Statistics}
\label{Cross}
\begin{tabular}{*{2}{@{\hskip 0.2in}c@{\hskip 0.2in}}} 
\toprule
\emph{Composition} & \emph{Cross Validation $\chi_{reduced}^2$} \\
\colrule
Zr$_{80}$Pt$_{20}$ & 0.0043631 \\
\\
Zr$_{70}$Pd$_{30}$ & 0.000973443 \\
\\
Ti$_{40}$Zr$_{10}$Cu$_{30}$Pd$_{20}$ & 0.00193925 \\
\\
Diopside & 0.097273  \\
\botrule
\end{tabular}
\end{table}
Overall, the various statistical measures and analyses employed to assess the validity and goodness of fit of the DEH model appear to objectively suggest that the functional form of Eq. (\ref{Final}), is able to accurately describe the phenomenology of the super-Arrhenius growth of the viscosity. This was demonstrated for 45 distinct and diverse supercooled liquids, and makes a strong case for the DEH as a descriptor of the glass transition.

\subsection{Statistical Comparison With Other Theories}
\label{the_others}

We have demonstrated both qualitatively and quantitatively, that the DEH model and functional form for the viscosity is able to accurately describe and reproduce the temperature dependence of the viscosity of supercooled liquids using only a \textbf{single} fitting parameter. It is important, however, to examine the DEH in comparison with existing theories and models of the glass transition, some of which even provide for collapse of the viscosity data \cite{BENK1}. To that end, we statistically compare and contrast the DEH with five of the most widely used models of supercooled liquids, and their associated functional forms for the viscosity. We selected a glass forming liquid from three of the classes considered, namely organic (OTP), silicate (LS2), and metallic ($Pd_{77.5}Cu_{6}Si_{16.5}$) and fit the viscosity functions arising from the KKZNT avoided critical point model \cite{KKZNT1,KKZNT2,KKZNT3},
\begin{eqnarray}
\label{KKZNT}
\ln\eta = \ln\eta_0 +\frac{E_{\infty}}{T}+\frac{T_{A}}{T}B[\frac{T_A-T}{T_A}]^z\Theta(T_A-T) 
\end{eqnarray}
Cohen-Grest free volume model \cite{CG},
\begin{eqnarray}
\label{CG}
\ln\eta=\ln\eta_0 +\frac{2B}{T-T_0+\sqrt{(T-T_0)^2+CT}} 
\end{eqnarray}
BENK modified parabolic model \cite{BENK1,BENK2,Blodgett}, 
\begin{eqnarray}
\label{BENK}
\ln\eta=\ln\eta_0 +\frac{E_{\infty}}{k_BT} +J^2(\frac{1}{T}-\frac{1}{\tilde{T}})^2 \Theta(\tilde{T}-T) 
\end{eqnarray}
MYEGA entropy model \cite{MYEGA},
\begin{eqnarray}
\label{MYEGA}
\ln\eta= \ln\eta_0 + \frac{K}{T}e^{\frac{C}{T}} 
\end{eqnarray}
and the oft-employed VFT form (Eq. \ref{VFT}). 

\begin{figure*}
	\centering
	\includegraphics[width=1.8 \columnwidth]{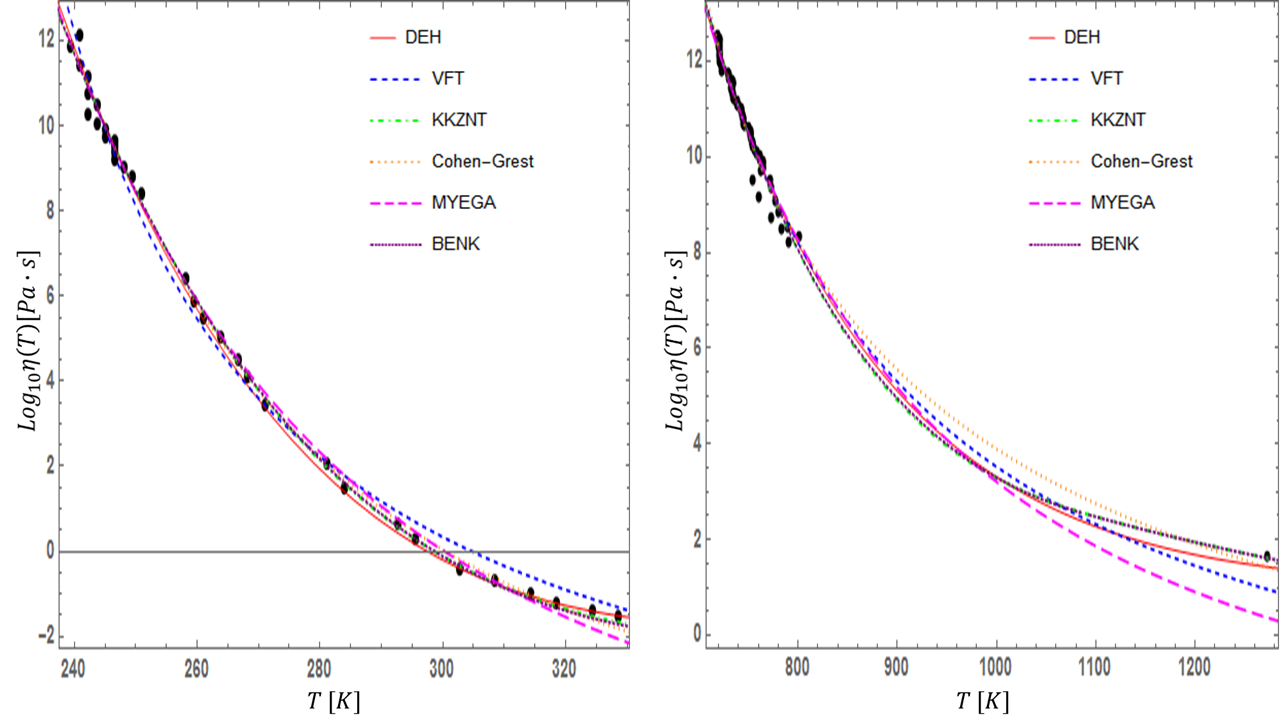}
	\caption{(Color Online.) Comparison of DEH [Eq. (\ref{Final})], VFT [Eq. (\ref{VFT})], KKZNT [Eq. (\ref{KKZNT})], Cohen-Grest [Eq. (\ref{CG})], MYEGA [Eq. (\ref{MYEGA})], and BENK [Eq. (\ref{BENK})] forms for the viscosity as applied to fragile OTP and strong LS2.}
	\label{OTPLS2Fits.}
\end{figure*}
\begin{figure*}
	\centering
	\includegraphics[width=1.8 \columnwidth, height= .3 \textheight]{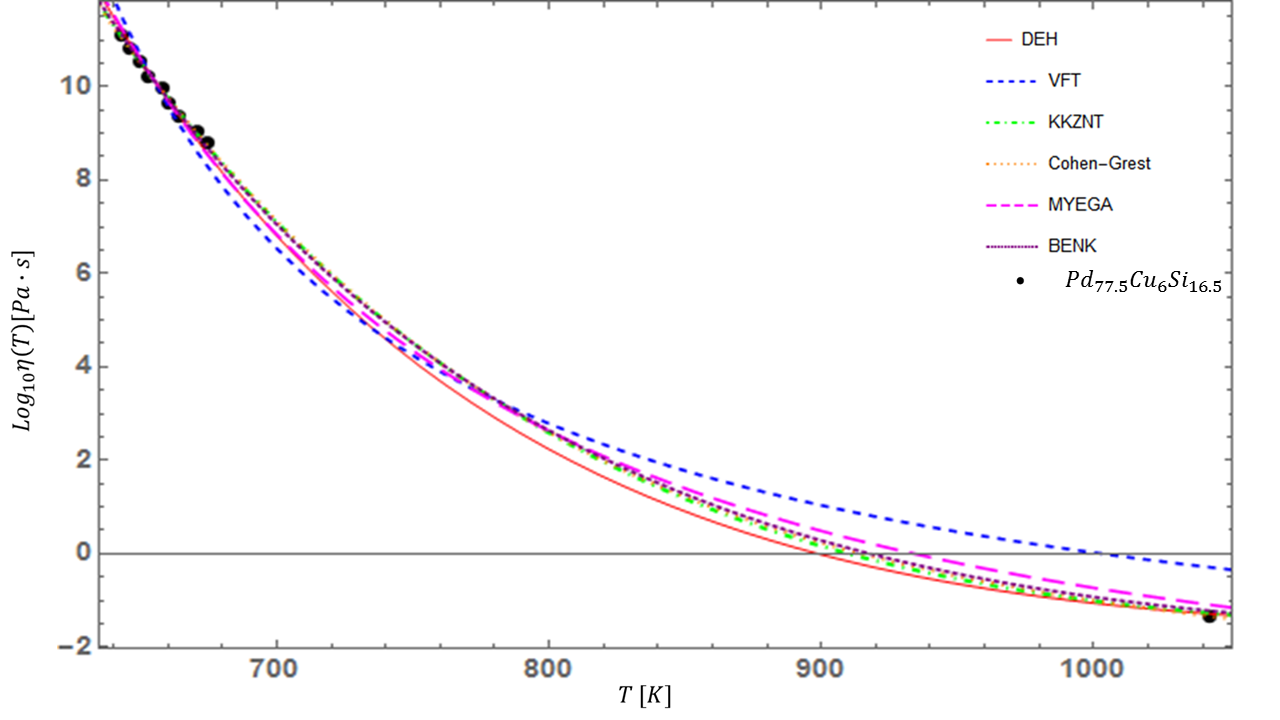}
	\caption{(Color Online.) Comparison of DEH [Eq. (\ref{Final})], VFT [Eq. (\ref{VFT})], KKZNT [Eq. (\ref{KKZNT})], Cohen-Grest [Eq. (\ref{CG})], MYEGA [Eq. (\ref{MYEGA})], and BENK [Eq. (\ref{BENK})] forms for the viscosity as applied to the metallic liquid $Pd_{77.5}Cu_{6}Si_{16.5}$.}
	\label{Metallic.}
\end{figure*}

\begin{figure*}
	\centering
	\includegraphics[width=1.8 \columnwidth]{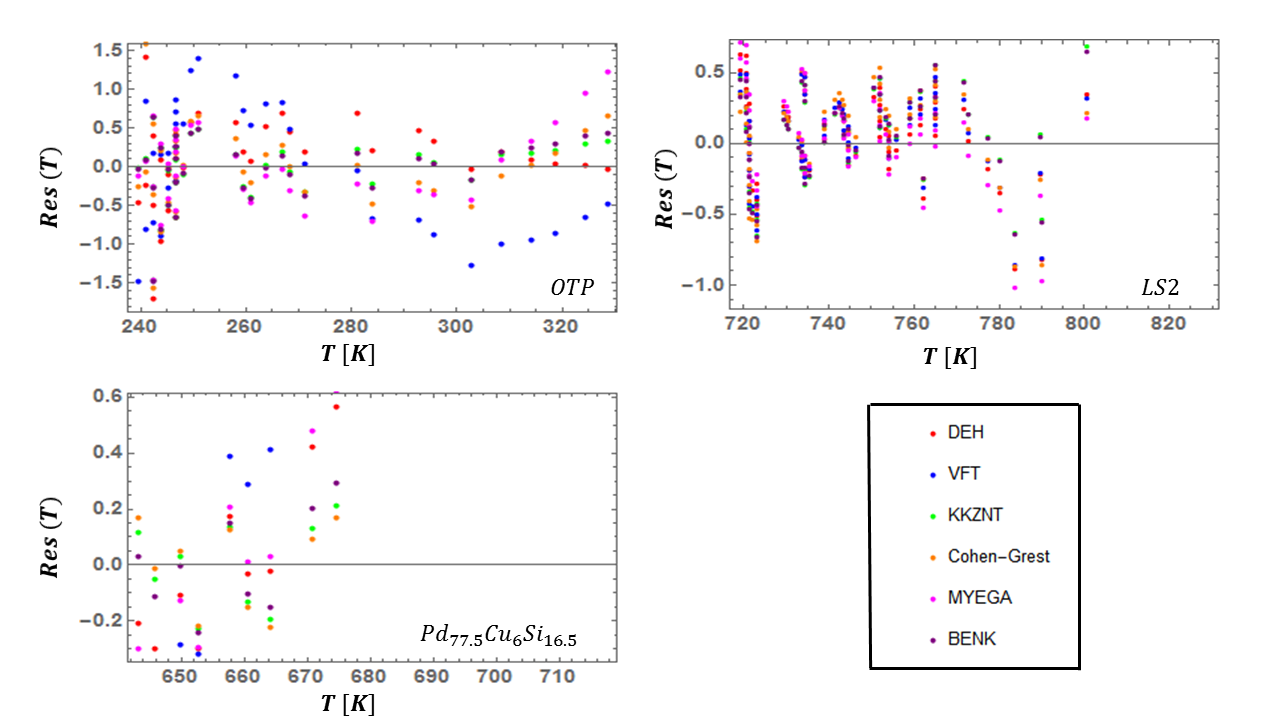}
	\caption{(Color Online.) Comparison of the residuals from Eq. (\ref{residual}) of the DEH [Eq. (\ref{Final})], VFT [Eq. (\ref{VFT})], KKZNT [Eq. (\ref{KKZNT})], Cohen-Grest [Eq. (\ref{CG})], MYEGA [Eq. (\ref{MYEGA})], and BENK [Eq. (\ref{BENK})] forms of the viscosity.}
	\label{Fit_Residuals.}
\end{figure*}

Each one of these functional forms (including VFT) has at least two parameters that (at this point) cannot be determined from first principles. In order to determine the parameters of the various forms for the three test liquids, we fit the natural logarithm of the viscosity data as a function of temperature. As the DEH form of Eq. (\ref{Final}) is applicable only below the liquidus temperature, $T_{l}$, we needed a `fair' metric for comparison between the above theories whose functional forms apply to the entire temperature range of measured data. To that end, in all cases, we first applied a linear fit to the high temperature Arrhenius (above the crossover temperature) regime as a function of inverse temperature. This allows us to extract values of the prefactor, $\eta_0$, and the extrapolated high temperature activation energy, $E_{\infty}$, where relevant. For each liquid, we fixed these values and then fit the various models over the remaining temperature range from $T_{l}$ and below, to extract the values of the remaining parameters. In the case of KKZNT, we also constrained the parameter $z$ to be $\frac{8}{3}$ (see \cite{KKZNT1} for discussion). The results of the fitting with the extracted parameters are depicted in Figures (\ref{OTPLS2Fits.}) and (\ref{Metallic.}). The functional forms of all studied theories are shown along with the DEH fits. In Figure (\ref{Fit_Residuals.}) we plot the residuals of all the forms together. A visual comparison of the fits against the data makes clear that the VFT form consistently provides the worst fit to the experimental data. Qualitatively, it is difficult to distinguish the goodness of fit of the remaining models. In order to resolve these differences, we examine the residuals of the fits in Fig. (\ref{Fit_Residuals.}) and compute statistical measures. The residuals are more or less consistent across the forms, with the exception of VFT, which as expected shows significant bias, especially in the case of OTP. This provides support to the argument that the non random nature of the residuals is likely related to measurement error in the data. The calculated statistical values are provided in Table (\ref{goodness}) in the rows corresponding to the various model designations. Examining the calculated values makes clear that the DEH model consistently outperforms both the VFT and MYEGA forms across all three liquids. The Cohen-Grest form appears to be roughly similar to the DEH form, whereas the KKZNT and BENK forms consistently outperform the DEH. This result is not surprising, as the KKZNT and BENK forms have many more open parameters than the DEH form. In fact, it can be shown that forcing the values of the ``special temperatures" that appear in these forms to correspond to experimental values causes the fit performance to worsen significantly. This is consistent with previous results that suggested that the KKZNT form does not perform well for silicate glassformers and that the MYEGA form does not perform well for metallic glass formers. This leads to the conclusion that in numerous cases, the optimal values of the parameters are often inconsistent with the underlying theoretical motivations. This suggests that the various models contrasted with the DEH are unable to universally describe the phenomenology of supercooled liquids of all types, unlike the DEH model.

\section{Universality Amongst Supercooled Liquids}
The DEH viscosity function of Eq. (\ref{Final}) has only one dimensionless parameter ($\bar{A}$). The two other quantities appearing in Eq. (\ref{Final}) (the liquidus temperature $T_{l}$ and the viscosity of the supercooled liquid at the liquidus temperature $\eta(T_{l})$) are both fixed by experiment. Thus, the DEH theory \cite{Nussinov} implies that the viscosity data of supercooled fluids may be collapsed onto one master curve with the judicious value of this single parameter $\bar{A}$. In \cite{ESDH}, we succinctly verified this prediction of a universality in the viscosity data. Here, we expand on this newfound universality.  In Section \ref{test.}, we demonstrated that the DEH model can accurately reproduce the viscosity of all types of supercooled liquids, over many decades. Thus, since Eq. (\ref{Final}) holds over the experimentally relevant temperature range, it follows the glass transition phenomenon possesses some form of universality that may be unearthed with the aid
of the single parameter $\bar{A}$. The virtue of the universality implied by Eq. (\ref{Final}) is that the only temperature and viscosity scales are the equilibrium liquidus temperature  
and the viscosity at this temperature. There are no assumptions about exotic temperatures such $T_{0}$ of Eq. (\ref{VFT}). Rather, the only pertinent temperature in the DEH framework is that associated with a known equilibrium transition. 
\begin{figure*}
\centering
\includegraphics[width=1.8 \columnwidth]{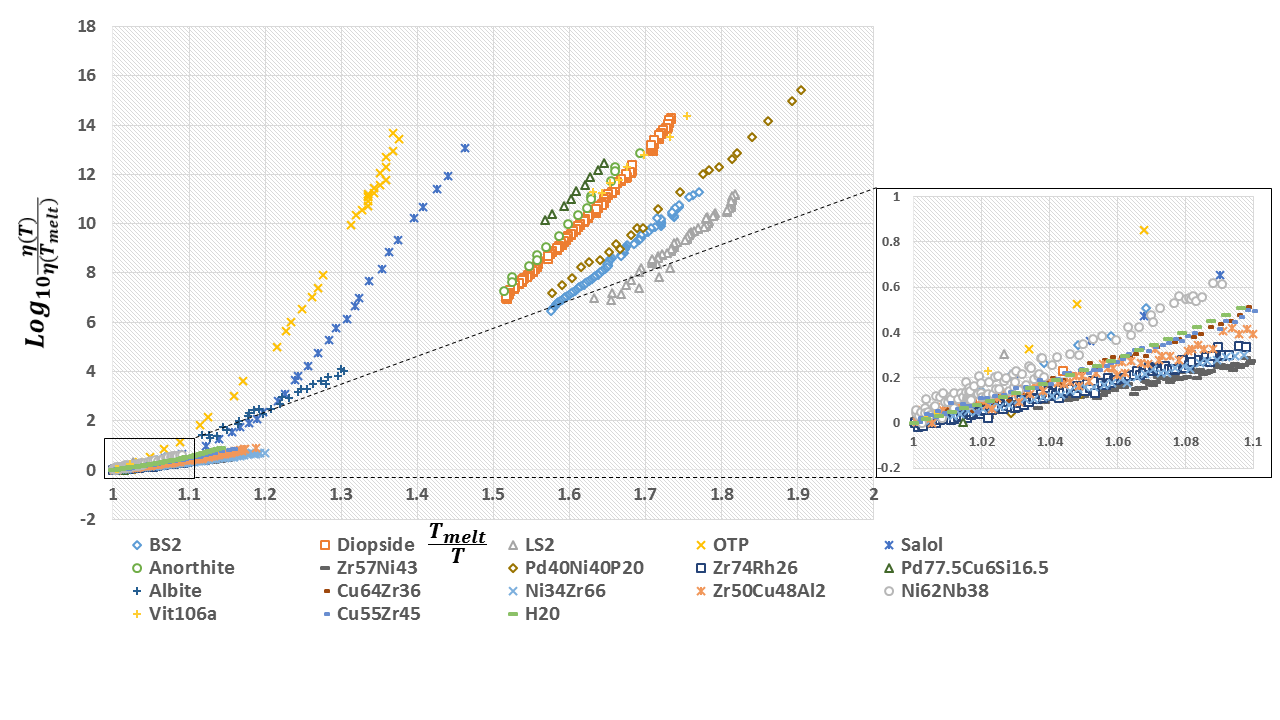}
\caption{(Color Online.) The logarithm of the viscosity, $\eta(T)$, scaled by the viscosity at the liquidus, $\eta(T_{l})$, versus the scaled temperature, $\frac{T_{l}}{T}$ for a subset of the studied liquids. When represented this way universal behavior does not appear, however, a spectrum of behavior approximating the fragility does appear. A careful inspection of the plot indicates that most glassformers seem to fall within different `families' corresponding to fragility classes as defined by experimental values. }
\label{Fragility.}
\end{figure*}

\begin{figure*}
\centering
\begin{subfigure}[t]{0.6\textwidth}
	\includegraphics[width=1 \columnwidth, height=.7 \textheight, keepaspectratio]{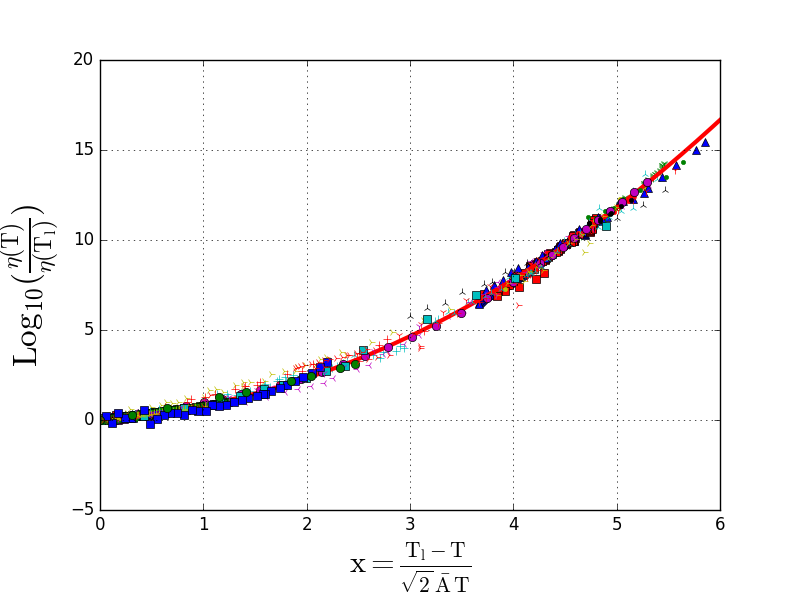}
	\caption{}\label{Collapse:Data}
\end{subfigure}
\begin{subfigure}[t]{0.35\textwidth}
	\includegraphics[width=1 \columnwidth, height=.7 \textheight, keepaspectratio]{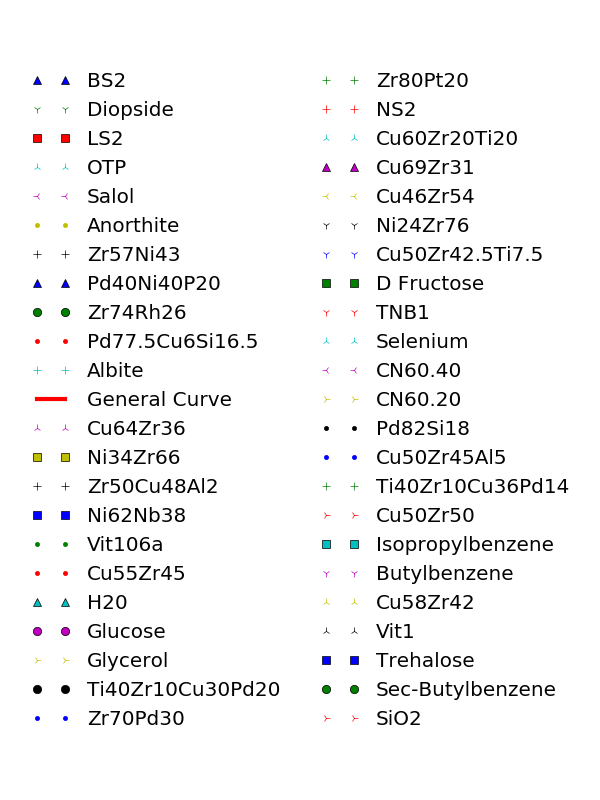}
	\caption{}\label{Collapse:Legend}
\end{subfigure}
\caption{(Color Online.) Log base-10 representation of the viscosity data scaled by its value at the liquidus, $\eta(T_{l})$, versus $x$, as defined in the figure for all studied liquids. The data of all 45 liquids, of all types  and kinetic fragilities, is observed to fall upon (`collapse onto') a single, universal scaling curve. This result is suggestive of an underlying universality in the behavior of supercooled liquids. Note the exceptional agreement over 16 decades.}
\label{Collapse.}
\end{figure*}

\begin{figure*}
\centering
\begin{subfigure}[t]{.45\textwidth}
	\centering
	\includegraphics[width=\linewidth]{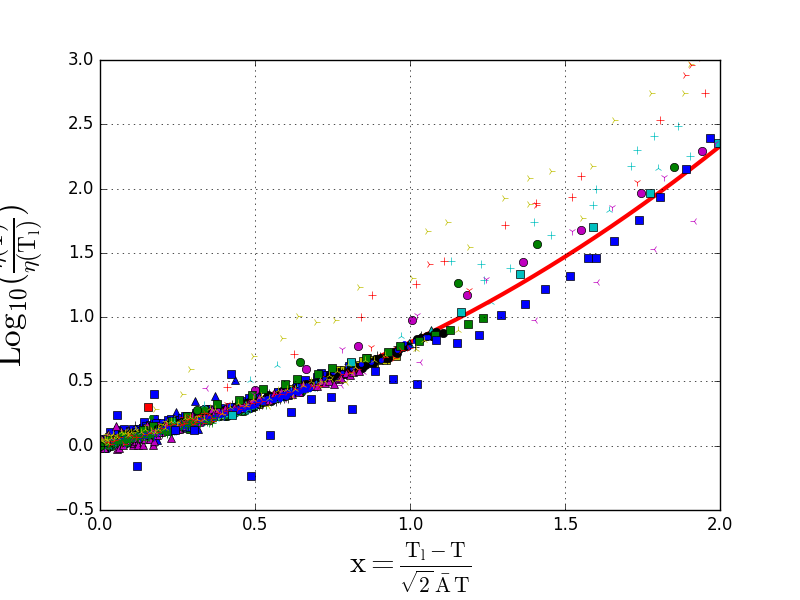}
	\caption{}\label{Zoom:Reg_1}
\end{subfigure}
\begin{subfigure}[t]{.45\textwidth}
	\centering
	\includegraphics[width=\linewidth]{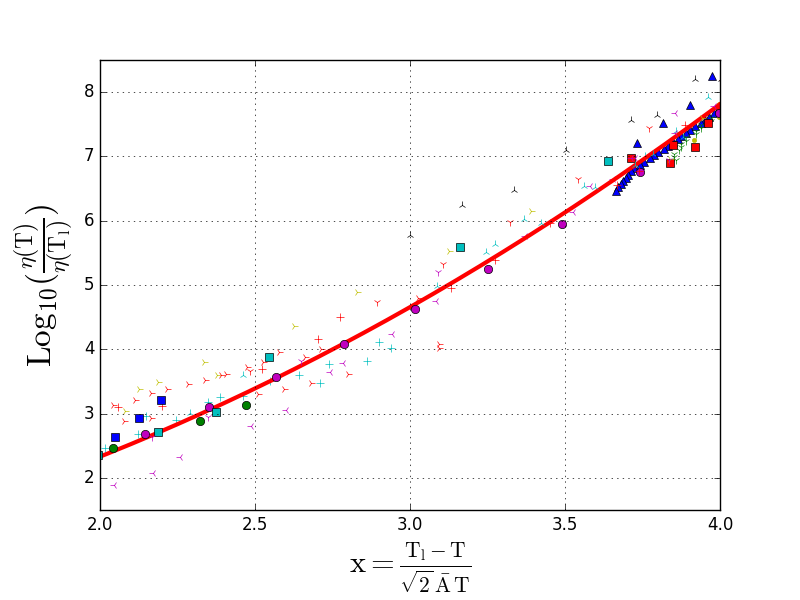}
	\caption{}\label{Zoom:Reg_2}
\end{subfigure}

\medskip

\begin{subfigure}[t]{.45\textwidth}
	\centering
	\vspace{0pt}
	\includegraphics[width=\linewidth]{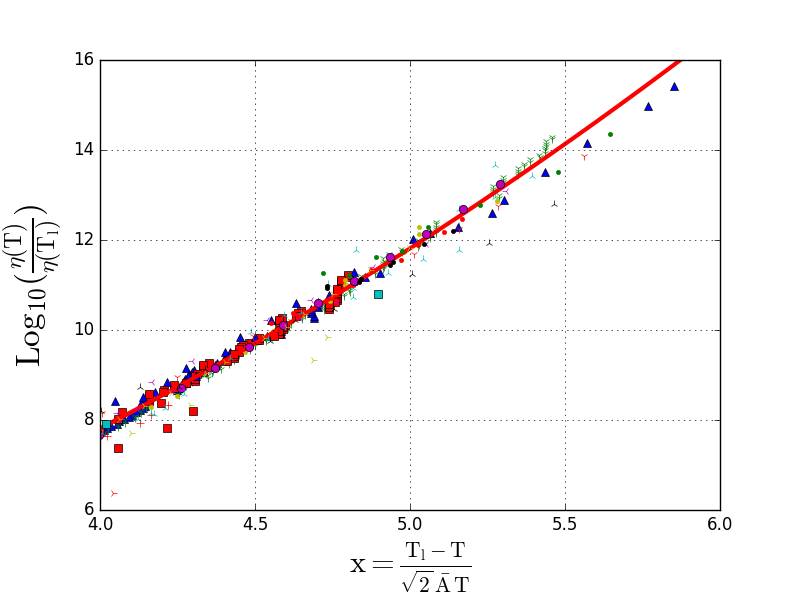}
	\caption{}\label{Zoom:Reg_3}
\end{subfigure}
\begin{minipage}[t]{.45\textwidth}
	\caption{(Color Online.) Zoomed in version of the viscosity collapse data presented in Fig.(\ref{Collapse.}). The data is presented in this way to demonstrate that the logarithmic form for the collapse is not masking or suppressing poor fits. Panel (\ref{Zoom:Reg_1}) focuses on the data in the immediate vicinity of the liquidus temperature. Panel (\ref{Zoom:Reg_2}) focuses on the mid temperature range. This region is the most sparse, as experimental data in this range is uncommon. Panel (\ref{Zoom:Reg_3}) shows the lowest temperature region, where the DEH fit is the tightest, and alternative models typically have poorest performance.}
\end{minipage}
\label{Zoom}
\end{figure*}
Previously, many investigations attempted to scale the viscosity data of all liquids studied in some meaningful way, to see if they can be collapsed onto a single curve
by appealing to various ``special" temperatures such as the just noted "ideal glass transition temperature'' $T_0$, the dynamic glass transition temperature $T_g$ (at which the viscosity reaches the rather arbitrary threshold value of $10^{12}$ Pascal $\times$ sec) that often coincides with a weak thermodynamic signature \cite{yyue}, or the Arrhenius cross-over temperature $T_A$  \cite{Blodgett,KKZNT1,KKZNT2,KKZNT3} above which the activated Arrhenius form of
Eq. (\ref{arrhenius}) holds and below which deviations from  Eq. (\ref{arrhenius}) fails and rigidity and collective phenomena emerge \cite{Soklaski,Soklaski1, Weingartner}. Within the DEH model, the equilibrium liquidus temperature $T_{l}$ constitutes the only temperature of significance. As we discussed above, Eq. (\ref{Final}) implies that a universal data collapse of the viscosity occurs relative to the
ratio $(T_{l}-T)/T$ when it is scaled by the single dimensionless parameter $\bar{A}$ of the DEH theory.

 In Fig \ref{Fragility.}, we plot the logarithm of the viscosity data as a function of the reciprocal temperature with both quantities scaled by their respective values at $T_{l}$. A moment's inspection reveals that this scaling does not lead to a data collapse. However, an interesting result does appear in the fact that the viscosity data appears to fall into a spectrum of ``fragility bands" in much the same way the standard Angell plot does. This suggests that fragilities appear quite naturally in the DEH model using $T_{l}$ (a well defined temperature) instead of $T_g$ (an occasionally arbitrarily defined temperature) as the fundamental scaling temperature. This is not unreasonable, as the Kauzmann ``2/3" rule \cite{232} suggests that on average $T_g=\frac{2}{3}T_{l}$. Therefore, we would expect that Angell-like fragility scaling should appear in the plot versus the reciprocal melting temperature, as it is, in many cases, proportional to the glass transition temperature. This will be discussed more in the next section (for a summary of earlier attempts to relate the glass transition to melting see, e.g., \cite{ESDH}).

Since the DEH model contains only a single fitting parameter, and is able to reproduce the viscosity of a disparate cross section of supercooled liquids using only this single parameter, it seems natural to include $\bar{A}$ in any scaling attempts. In Fig. \ref{Collapse.} we again plot the logarithm of the scaled viscosity, $\eta(T)/\eta(T_{l})$, but this time against the argument of the complementary error function in the DEH form for the viscosity (Eq. \ref{Final}), namely $x \equiv \frac{T_{l} -T}{T \bar{A} \sqrt{2}}$. It is immediately apparent that this scaling collapses the viscosities of the 45 liquids shown in this figure onto a \textbf{single, universal curve} over as much as \textit{16} decades (the gray curve in the figure represents the theoretical prediction expected in the DEH \cite{Nussinov}, Eq. \ref{Final}). The collapse in Fig. \ref{Collapse.} (originally reported in \cite{ESDH}), to our knowledge represents one of the first times that data for all types of supercooled liquids has been collapsed over this many orders of magnitude. This result suggests that the mechanism of the glass transition is, in fact, universal across all liquids, and warrants significant further investigation. In the preceding section we undertook a rigorous statistical analysis to objectively assess the performance of the DEH model. The statistical analysis performed on the individual compositions is equivalent to performing a statistical analysis of the goodness of fit for the universal collapse curve, so a separate analysis of the universal curve itself will not be undertaken. However, in Figure \ref{Zoom:Reg_1}-\ref{Zoom:Reg_3} we ``zoom in" on the three major regimes in the universal curve, namely temperatures very near melting, midrange temperatures, and temperatures corresponding to deeply supercooled liquids where glassy effects are strongest. The objective of this ``zooming" is to demonstrate that we are not ``sweeping anything under the rug", so-to-speak, by representing the data in this way. It is standard to plot the logarithm of viscosity data when demonstrating collapse-like curves, but in the interest of total transparency, we show that even at higher ``resolution" the data fits the curve of the DEH with only minor spread. From the figure it is clear that in the deeply supercooled region, the collapse is very tight, with little visible spread. Indeed, in the two regions immediately beneath melting, the data also falls along the universal curve with minimal spread, but what spread does exist, is likely due to the influence of the ${\cal{PT}}E{\cal{I}}$ region discussed at length above. With this visual understanding of the accuracy of the DEH model in focus, we now move on to make objective, quantitative measures of said accuracy. 

\section{Temperature Dependence of $\bar{\sigma}_{T}$}
\begin{figure*}
\centering
\includegraphics[width=1.8 \columnwidth]{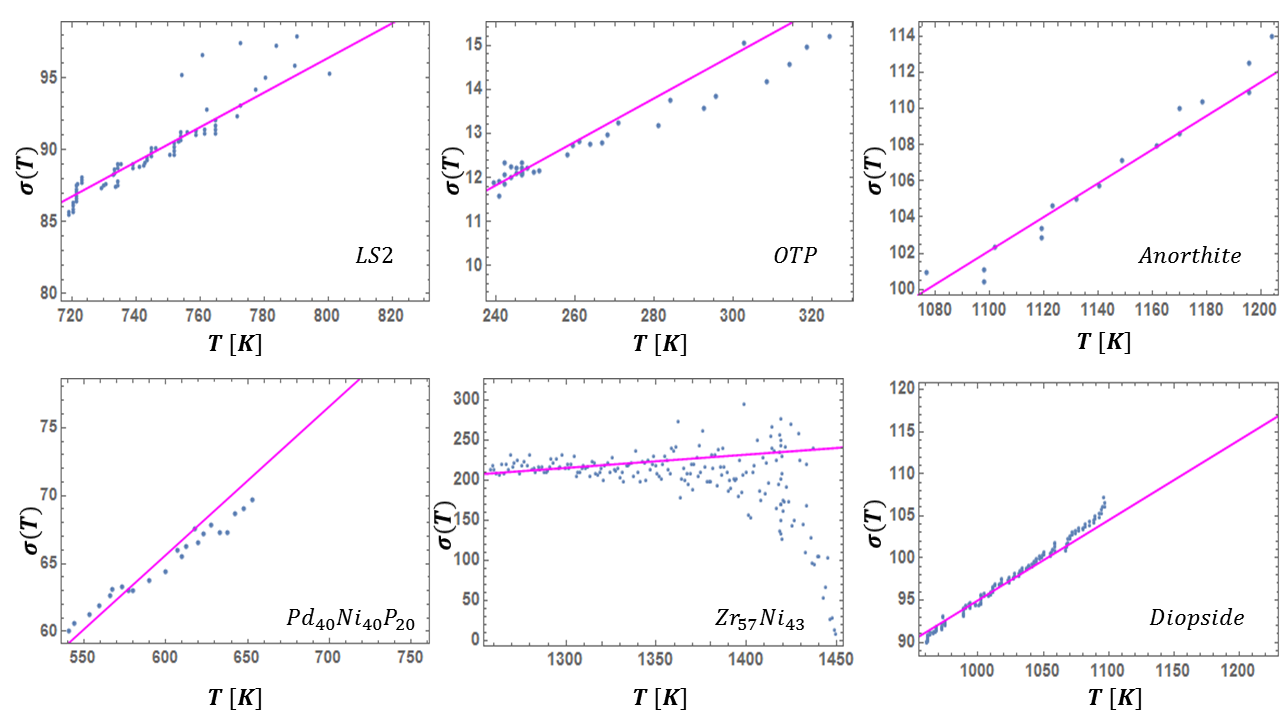}
\caption{(Color Online.) Plot of $\bar{\sigma}_{T}$ as a function of $T$, found by inverting the expression in Eq.(\ref{Final}) and applying it to experimental data. The fit from the linear assumption is shown to work very well for most cases, especially at temperatures far below melting.}
\label{Sigma.}
\end{figure*}

\begin{figure*}
\centering
\includegraphics[width=1.8 \columnwidth]{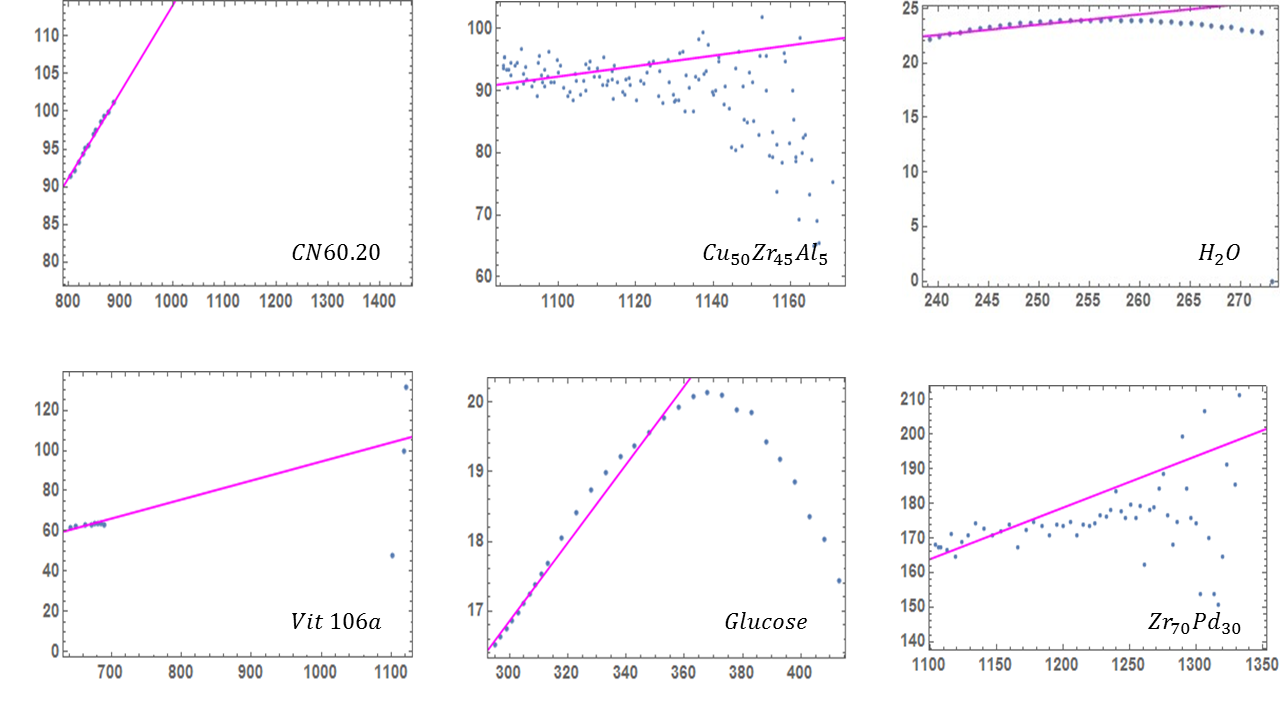}
\caption{(Color Online). Additional examples of $\bar{\sigma}_{T}$ as a function of $T$. The linear fit well approximates most systems over much of the temperature range. Glucose displays anomalous behavior which is discussed in the Supplementary Information (\ref{sec:anomaly}).}
\label{Sigma2.}
\end{figure*}

A major assumption of the DEH model is that the width, $\bar{\sigma}_{T}$, of the distribution of eigenstates has an approximately linear temperature dependence, i.e. $\frac{d\bar{\sigma}}{dT}=const\equiv \bar{A}$. As this notion is fundamental to the form of the viscosity function applied throughout this paper, it is imperative that the validity of this assumption be checked. Assuming that all other facets of the DEH model up to the assumption of the temperature dependence of $\bar{\sigma}$ is correct, Eq. (\ref{Final}) can be inverted to solve for the temperature dependence of $\bar{\sigma}_{T}$ using the experimental viscosity data. The results of this analysis for 12 various example liquids is presented in Figures (\ref{Sigma.}) and (\ref{Sigma2.}). Examining the results in the figures makes clear that in most cases the assumption of linearity of the width is accurate over a wide range of temperatures. In some of the cases, such as many of the metallics, the seeming bend as $T_{l}$ is approached is likely due to the precarious limit that arises as the temperature approaches $T_{l}$ from below in the inverted expression for $\bar{\sigma}_{T}$. There are also cases where a clear crossover in the behavior of $\bar{\sigma}_{T}$ takes place at various temperatures. These anomalies may be linked to various kinetic crossovers or hypothesized phase transitions. Although the linear assumption is generally valid, in some cases it appears as though a nonzero intercept may improve the fit quality. This would introduce a second parameter, and only make marginal improvements in the fit quality. For these and other reasons, a zero intercept is assumed for all liquids.


\section{The Physical Content of the single DEH parameter $\bar{A}$}
Having established in previous sections that the DEH model ably describes the behavior of supercooled liquids of all types with only a \textbf{single} fitting parameter, it is prudent to attempt to understand the physical meaning of $\bar{A}$. By definition, $\bar{A}$ governs the temperature dependence of the width of the distribution of eigenstates, which underlies the metastable features of the supercooled liquid. However, it is quite possible that it is linked to various thermodynamic or kinetic properties of the system. Uncovering any correlations between $\bar{A}$ and other physical quantities may help in increasing the understanding of both the DEH model and the glass transition itself. Additionally, finding a link between $\bar{A}$ and other measurable quantities may allow for the prediction of $\bar{A}$ and reduce the DEH model to a zero parameter theory. In what follows, we examine the relationship between $\bar{A}$ and certain physical observables, and find a number of interesting correlations.
\subsection{Fragility.}
\begin{figure}
\centering
\includegraphics[width=1 \columnwidth, height=.35 \textheight, keepaspectratio]{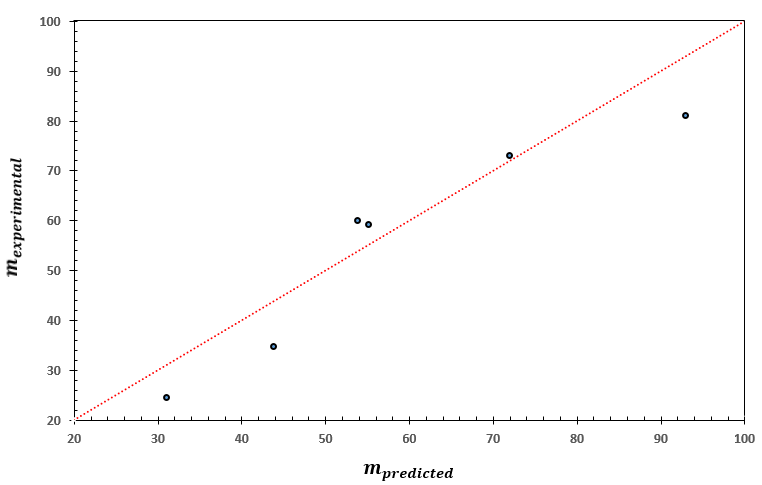}
\caption{(Color Online.) Experimental values of the fragility, $m_{measured}$, versus the values computed ($m_{DEH}$) using Eq. (\ref{fragility}). Overall, the correlation is strong, suggesting broad agreement between the DEH and experiment. However, the slope of the dashed line is not exactly equal to one, indicating a prefactor may be necessary, despite the strong correlation.}
\label{Fragility2.}
\end{figure}
\begin{table}
\centering
\caption{Relevant Temperatures}
\label{Temps}
\begin{tabular}{*{5}{@{\hskip 0.1in}c@{\hskip 0.1in}}}
\toprule
\emph{Composition} & \emph{$T_{l}$ {[}K{]}} & \emph{$T_g$ [K]} & \emph{$m_{measured}$} & \emph{$m_{DEH}$}  \\ 
\colrule
BS2 & 1699 & 973 & N/A & 47  \\
Diopside & 1664 & 933 & 59 & 68 \\
LS2 & 1307 & 727 & 35 & 44 \\
OTP & 329 & 240 & 81 & 93 \\
Salol & 315 & 220 & 76 & 72 \\
Anorthite & 1823 & 1113 & 69 & 54 \\
Zr$_{57}$Ni$_{43}$ & 1450 & 722 & N/A & 33  \\
Pd$_{40}$Ni$_{40}$P$_{20}$ & 1030 & 560 & 50 & 57  \\
Zr$_{74}$Rh$_{26}$ & 1350 & 729 & N/A & 40  \\
Pd$_{77.5}$Cu$_6$Si$_{16.5}$ & 1058 & 632 & 64 & 64  \\
Albite & 1393 & 1087 & 24 & 31 \\
Cu$_{64}$Zr$_{36}$ & 1230 & 800 & N/A & 36 \\
Ni$_{34}$Zr$_{66}$ & 1283 & 760 & N/A & 24 \\
Zr$_{50}$Cu$_{48}$Al$_{2}$ & 1220 & 675 & N/A & 46 \\
Ni$_{62}$Nb$_{38}$ & 1483 & N/A & N/A & N/A \\
Vit106a & 1125 & 672 & 45 & 56 \\
Cu$_{55}$Zr$_{45}$ & 1193 & N/A & N/A & N/A \\
H$_2$O & 273 & 136 & N/A & 100 \\
Glucose & 419 & N/A & 72 & N/A \\
Glycerol & 291 & 190 & 53 & 61 \\
Ti$_{40}$Zr$_{10}$Cu$_{30}$Pd$_{20}$ & 1280 & 687 & N/A & 41 \\
Zr$_{70}$Pd$_{30}$ & 1351 & N/A & N/A & N/A \\
Zr$_{80}$Pt$_{20}$ & 1364 & N/A & 45 & N/A \\
NS2 & 1147 & N/A & N/A & N/A \\
Cu$_{60}$Zr$_{20}$Ti$_{20}$ & 1125 & 647 & N/A & 106 \\
Cu$_{69}$Zr$_{31}$ & 1313 & N/A & N/A & N/A \\
Cu$_{46}$Zr$_{54}$ & 1198 & N/A & N/A & N/A \\
Ni$_{24}$Zr$_{76}$ & 1233 & 626 & N/A & N/A \\
Cu$_{50}$Zr$_{42.5}$Ti$_{7.5}$ & 1152 & 677 & N/A & 48 \\
D Fructose & 418 & N/A & N/A & N/A \\
NB1 & 472 & N/A & N/A & N/A \\
Selenium & 494 & 308 & 87 & 50 \\
CN60.40 & 1170 & N/A & N/A & N/A \\
CN60.20 & 1450 & N/A & N/A & N/A \\
Pd$_{82}$Si$_{18}$ & 1071 & N/A & 106 & N/A \\
Cu$_{50}$Zr$_{45}$Al$_{5}$ & 1173 & 701 & N/A & 71 \\
Ti$_{40}$Zr$_{10}$Cu$_{36}$Pd$_{14}$ & 1185 & 669 & N/A & 64 \\
Cu$_{50}$Zr$_{50}$ & 1226 & 651 & 60 & 53 \\
Isopropylbenzene & 177 & 126 & 74 & 94 \\
Butylbenzene & 185 & 128 & 60 & 79 \\
Cu$_{58}$Zr$_{42}$ & 1199 & N/A & N/A & N/A \\
Vit 1 & 937 & 625 & 54 & 46 \\
Trehalose & 473 & 380 & N/A & 54 \\
Sec-Butylbenzene & 190.3 & 127 & N/A & 102 \\
SiO$_2$ & 1873 & 1475 & 20 & 38 \\
\botrule
\end{tabular}
\end{table}
It is almost universally accepted that the concept of fragility in glass science is intimately linked to dynamics and structure in supercooled liquids, and not merely an artificial scaling property. Indeed, recent studies have shown that fragility can be related to the temperature dependence of the structure factors, and radial distribution functions upon supercooling, and therefore represents a measurable physical quantity of significance to glass theory \cite{Kelton}. Therefore, we must briefly examine the nature of fragility in the DEH. We saw in Section III D that a fragility spectrum similar to Angell's classic plot \cite{Angell} appeared quite naturally when scaling the viscosity and reciprocal temperatures about melting. We can take this further by using the definition of the fragility parameter, $m$, as outlined in \cite{Angell}. Using this definition and the functional form of the viscosity in the DEH (\ref{Final}), an explicit expression for the fragility can be derived within the DEH framework as
\begin{eqnarray}
\label{fragility}
\left.m\equiv \frac{d\log_{10}\eta(T)}{d(T_{g}/T)}\right|_{T=T_g} = {\frac{\sqrt{\frac{2}{\pi}}}{\ln(10)}} {\frac{T_{l}}{T_{g}}}{\frac{1}{\bar{A}}} {\frac{e^{-\left({\frac{\frac{T_{l}}{T_g}-1}{\sqrt{2} \bar{A}}}\right)^2}}{{\rm erfc}\left[{\frac{\frac{T_{l}}{T_g}-1}{\sqrt{2} \bar{A}}}\right]}}.
\end{eqnarray}
Strikingly, and perhaps surprisingly, the reduced temperature from the Turnbull criterion \cite{Turnbull,Turnbull1} for glass forming ability (GFA), $T_{red} \equiv \frac{T_g}{T_{l}}$, appears throughout this expression. This provides further support to a link between fragility and GFA. We can calculate the values of the fragility using Eq. (\ref{fragility}) and compare with the experimental values reported in \cite{Fragility,Blodgett}. The results of this comparison are presented in Table (\ref{Temps}) and in Figure (\ref{Fragility2.}). From Fig. (\ref{Fragility2.}), it is clear that there is a correlation between the predicted and experimental values of the fragility. The slope of the line of fit is not exactly equal to unity, but this may be related to differences in the measurement of $T_g$. 

\subsection{The Crossover Temperature $T_A$ and Structural Considerations}
\label{sec:cross}
\begin{figure*}
\centering
\includegraphics[width= 1.8 \columnwidth]{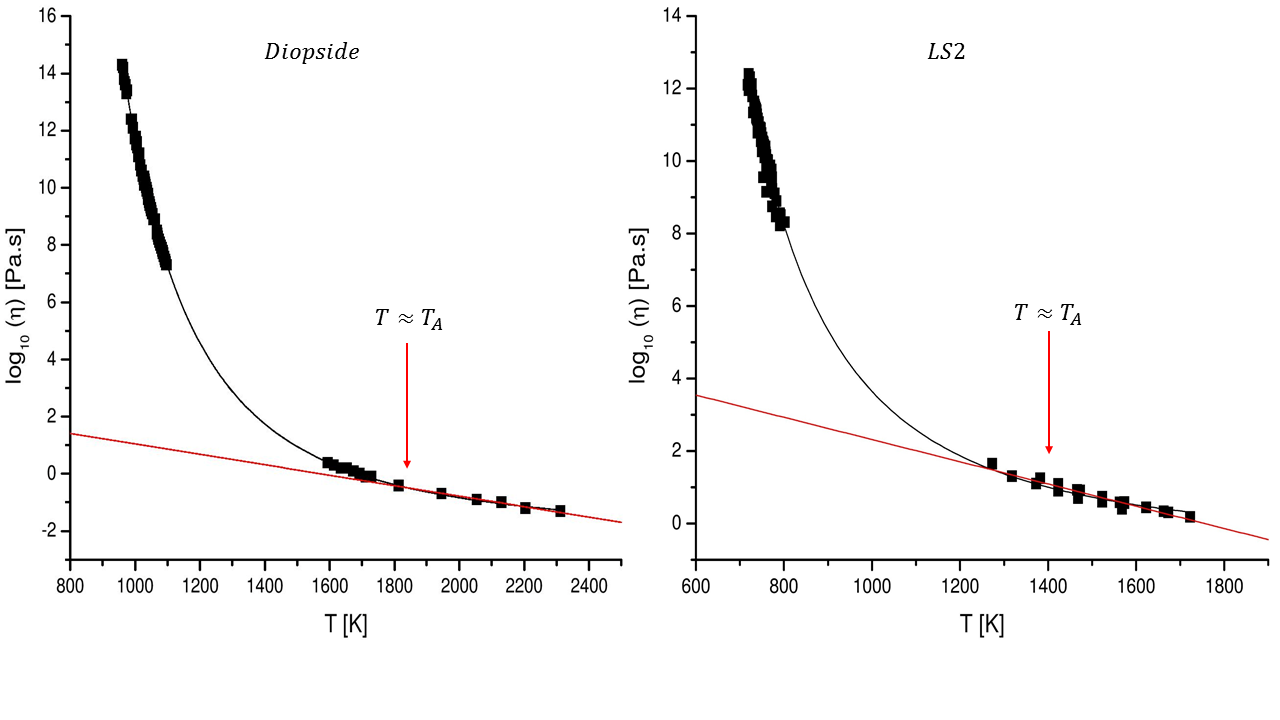}
\caption{(Color Online). Log base-10 viscosity Data as a function of temperature for Diopside and LS2. The changeover from Arrhenius behavior appears to occur in the vicinity of 1900 K for diopside and 1450-1500 K for LS2, roughly agreeing with predicted values for the crossover temperature $T_A$.}
\label{CompTa.}
\end{figure*}
\begin{table}
\centering
\caption{Predicted Values of Crossover Temperature, $T_A$}
\label{Ta}
\begin{tabular}{*{2}{@{\hskip 0.5in}c@{\hskip 0.5in}}}
\toprule
\emph{Composition} & \emph{$T_{A}$ {[}K{]}} \\ 
\colrule
BS2 & 1911 \\
Diopside & 1839 \\
LS2 & 1486 \\
OTP & 346 \\
Salol & 336 \\
Anorthite & 2010 \\
Zr$_{57}$Ni$_{43}$ & 1738 \\
Pd$_{40}$Ni$_{40}$P$_{20}$ & 1157 \\
Zr$_{74}$Rh$_{26}$ & 1557 \\
Pd$_{77.5}$Cu$_6$Si$_{16.5}$ & 1160 \\
Albite & 1503 \\
Cu$_{64}$Zr$_{36}$ & 1368 \\
Ni$_{34}$Zr$_{66}$ & 1506 \\
Zr$_{50}$Cu$_{48}$Al$_{2}$ & 1384 \\
Ni$_{62}$Nb$_{38}$ & 1607 \\
Vit106a & 1242 \\
Cu$_{55}$Zr$_{45}$ & 1329 \\
H$_2$O & 302 \\
Glucose & 444 \\
Glycerol & 315 \\
Ti$_{40}$Zr$_{10}$Cu$_{30}$Pd$_{20}$ & 1472 \\
Zr$_{70}$Pd$_{30}$ & 1587 \\
Zr$_{80}$Pt$_{20}$ & 1549 \\
NS2 & 1268 \\
Cu$_{60}$Zr$_{20}$Ti$_{20}$ & 1214 \\
Cu$_{69}$Zr$_{31}$ & 1478 \\
Cu$_{46}$Z$r_{54}$ & 1348 \\
Ni$_{24}$Zr$_{76}$ & 1491 \\
Cu$_{50}$Z$r_{42.5}$Ti$_{7.5}$ & 1287 \\
D Fructose & 433 \\
TNB1 & 499 \\
Selenium & 544 \\
CN60.40 & 1308 \\
CN60.20 & 1637 \\
Pd$_{82}$Si$_{18}$ & 1186 \\
Cu$_{50}$Zr$_{45}$Al$_{5}$ & 1280 \\
Ti$_{40}$Zr$_{10}$Cu$_{36}$Pd$_{14}$ & 1313 \\
Cu$_{50}$Zr$_{50}$ & 1390 \\
Isopropylbenzene & 187 \\
Butylbenzene & 197 \\
Cu$_{58}$Zr$_{42}$ & 1322 \\
Vit 1 & 1017 \\
Trehalose & 498 \\
Sec-Butylbenzene & 202 \\
SiO$_2$ & 2002 \\
\botrule
\end{tabular}
\end{table}

\begin{figure}
\includegraphics[width=1 \columnwidth, height= .25 \textheight, keepaspectratio]{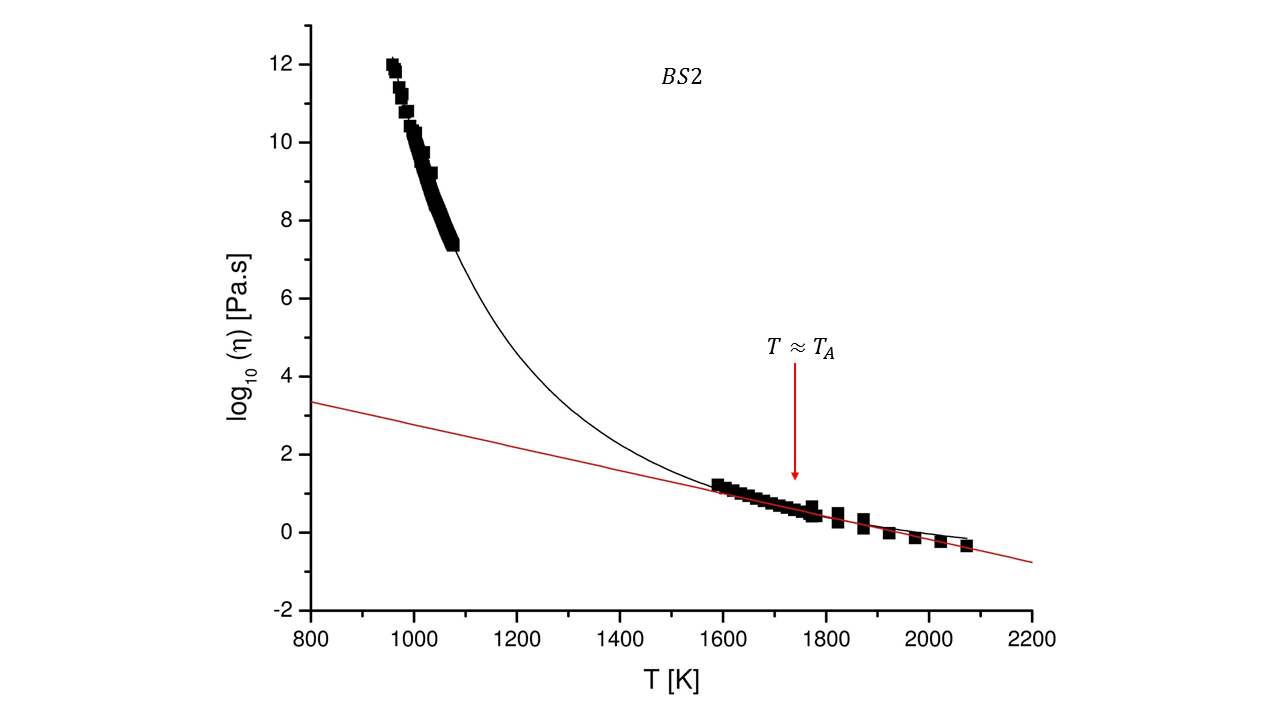}
\caption{(Color Online.) Log base-10 viscosity data as a function of temperature for BS2. The changeover from Arrhenius behavior appears to occur in the vicinity of 1750 K. The predicted crossover temperature $T_A$ was 1911 K. This agrees to within roughly ten percent.}
\label{BS2_Ta.}
\end{figure}

In the previous section we discussed the behavior of the fragility in the DEH framework. The fragility represents the degree of departure from Arrhenius behavior of the viscosity, and this deviation sets in at a crossover temperature, commonly known as $T_A$. It is natural to consider what meaning this crossover temperature will have within the DEH model, as the fragility appeared quite naturally in this framework. It is worth briefly outlining the general phenomenology of the crossover temperature, $T_A$, before proceeding. $T_A$ marks the temperature at which the super-Arrhenius growth of the viscosity sets in, the Stokes-Einstein relation breaks down, phonons delocalize, and cooperative motion of atoms/molecules/polymer chains in the liquid first begins \cite{Cavagna,Egami,Soklaski,Soklaski1}. The onset of the above phenomenon has been shown, in molecular dynamics simulations, to be correlated with the onset of structural changes associated with the formation, and subsequent percolation, of locally-preferred structures, which may or may not be subunits of the low temperature crystalline order \cite{Soklaski,Soklaski1,Weingartner}. All of these features are associated with the metastable, supercooled liquid. It seems reasonable, then, that upon cooling, the temperature at which these phenomena onset corresponds to the temperature at which the eigenstate distribution first has appreciable weight in the solid-like eigenstates \cite{Nussinov}. At high enough temperatures, even a distribution of nonzero width will not have a tail that has weight in the solid-like states, so flow should not be uninhibited, and structures in the supercooled liquid should more-or-less be consistent with structures present in the equilibrium liquid. It is clear that as temperature is lowered, the distribution widens, and shifts to lower energy states. At a certain temperature, the distribution over eigenstates will begin to have appreciable weight in the solid-like states. Structural and dynamical properties of these solid-like states should then start to play a role in the behavior of the liquid. We postulate that the temperature at which this occurs should be identified as the crossover temperature, $T_A$. With this idea in mind, we can make a rough approximation as to how $T_A$ should relate to the distribution in the DEH model. Assuming that at $T_A$ the width first spans the `distance' between the crossover temperature and the liquidus temperature, $T_{l}$ so as to have weight in the solid-like states, we can conjecture that at $T_{A}$ the 
 $\bar{\sigma}_{T}$, namely, $\bar{\sigma}_{T}=\bar{A}T$, is such that it extends up to energy density at the melting temperature $T_{l}$, i.e.,  
\begin{eqnarray}
\label{Ta1}
\bar{\sigma}_{T_A} \simeq \bar{A}( T_A-T_{l}).
\end{eqnarray}
This then yields \cite{Nussinov},
\begin{eqnarray}
\label{Ta2}
T_A \simeq \frac{T_{l}}{1-\bar{A}}.
\end{eqnarray}
Eq. (\ref{Ta2}) constitutes a rough approximation. The notion of 'appreciable weight' is rather arbitrary and there may be a prefactor that is necessary, and perhaps material dependent, in Eq. (\ref{Ta1}) (this will be investigated in an upcoming work). Despite this, we can examine the predicted values for $T_A$. The values of $T_A$ that are predicted by Eq. (\ref{Ta2}) are listed in the Table (\ref{Ta}). Previous studies of OTP and Pd$_{40}$Ni$_{40}$P$_{20}$ found $T_A$ values of approximately 350 K and 1157 K, respectively, which is in exceptional agreement with the values predicted from our approximation. Approximating $T_A$ from measured viscosity data is achieved by finding the temperature at which the viscosity ceases to be Arrhenius. In Fig. (\ref{CompTa.}) we display the viscosity data for both Diopside and LS2. Visually, it appears that deviations from Arrhenius behavior onsets at about 1839 K and 1486 K, respectively. These values are in good agreement with those predicted from our approximation. It is worth noting, however, that Eq. (\ref{Ta2}) does not apply for all 45 liquids studied. Eq. (\ref{Ta2}) predicts a value of 1911 K for BS2, which is clearly too high in comparison to the crossover point in Fig. (\ref{BS2_Ta.}). A discrepancy in values was also observed in the case of Pd$_{77.5}$Cu$_{6}$Si$_{16.5}$. A more detailed investigation of the exact relation between $\bar{A}$ and $T_A$ is clearly necessary yet preliminary results are promising.

The possible link between $\bar{A}$ and $T_A$ is significant not only because of the implications to the physics of the DEH model, but also because it may predict $\bar{A}$. Subsequently, this will enable the prediction of the viscosity in the entirety of the low temperature range, solely from high temperature data. This idea warrants more investigation, and further considerations for predicting $\bar{A}$ from high temperature data are discussed in Section \ref{sec:anomaly}.

It makes physical sense that having substantial weight of solid-like states for $T \leq T_A$ corresponds to the formation of locally preferred solid structures. The low-temperature locally preferred atomic structures would either be inherent to the equilibrium crystalline eigenstates, or result from the spatial mismatch of multiple crystalline ordered states in the distribution. In fact, the macroscopically disordered atomic arrangement of the glass logically results from overlap of multiple crystalline states of differing phonon modes and structural excitations. One can imagine cutting and superposing patches of elastically deformed lattice structures with given phonon modes associated with the relevant distribution of eigenstates. This would lead to the emergence of an overall `amorphous' structural arrangement. At this stage, this is an unproven conjecture. As we alluded to above, numerous studies have shown that various crystal-like or polyhedral structures begin to form at this temperature, $T_A$ . e.g., \cite{Soklaski,Soklaski1,Weingartner}. Further arguments based on uncertainties bolster these conclusions \cite{Nussinov}.
\begin{figure*}
\centering
\includegraphics[width=2 \columnwidth]{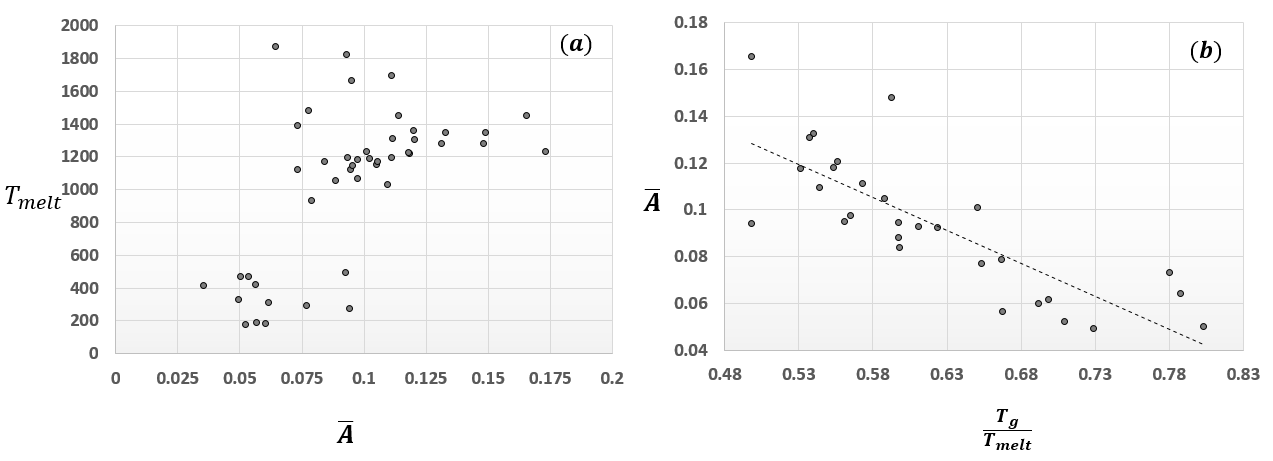}
\caption{(Color Online.) Correlation of $\bar{A}$ with various physical quantities: (a) $T_{l}$ versus $\bar{A}$ and (b) $\bar{A}$ versus Turnbull's reduced temperature, $T_r \equiv \frac{T_{l}}{T_g}$. $\bar{A}$ is seen to have a strong correlation with $T_r$, possessing a correlation coefficient of $r \approx 0.88$.}
\label{Corr.}
\end{figure*}
\begin{figure*}
\centering
\includegraphics[width=2 \columnwidth]{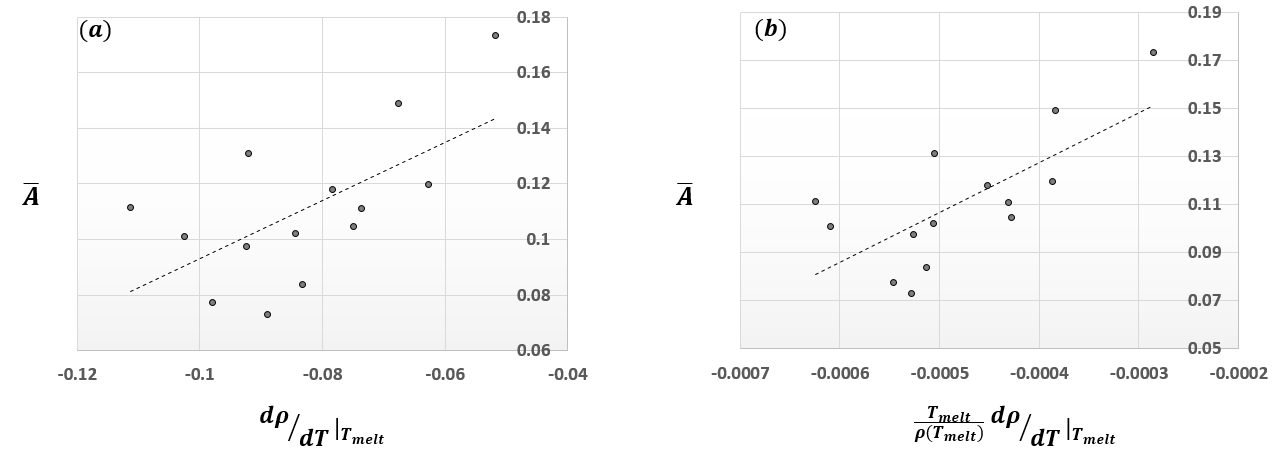}
\caption{(Color Online.) Correlation of $\bar{A}$ with the rate of change of density at melting in both the (a) bare and (b) scaled cases. A positive correlation is apparent, which can be rationalized in the DEH framework.}
\label{Corr2.}
\end{figure*}

\subsection{Correlation of $\bar{A}$ with Various Physical Quantities}
\label{Corr}
It has been demonstrated that the parameter, $\bar{A}$, appears in expressions for the fragility and crossover temperature within the DEH framework. Additionally, it is seemingly apparent from the raw values of $\bar{A}$ presented in Table (\ref{A}) that trends in the values may exist for similar liquid types. In light of these observations, it is reasonable to hypothesize that $\bar{A}$ may be linked to other macroscopic kinetic and thermodynamic properties of the supercooled liquid and glassy state. Toward this end, we investigated possible relationships between $\bar{A}$ and various physical properties that are relevant to the glass transition phenomenology. We begin by examining the possible relationship between $\bar{A}$ and two of the dynamical characteristics of glass forming liquids, namely the kinetic fragility parameter, $m$ (discussed in much more detail in Section VA), and the glass transition temperature, $T_g$. No direct discernible correlation was found with either quantity. This is not so surprising, as the exact expression for the kinetic fragility which was derived in Section V:A contained multiple other factors. Turning now to thermodynamic variables, we examine the relationship between $\bar{A}$ and the liquidus temperature, $T_{l}$ and the reduced glass transition temperature, $T_r \equiv \frac{T_{l}}{T_g}$, defined by Turnbull \cite{Turnbull}. Panel (a) of Fig. (\ref{Corr.}) shows the behavior of $T_{l}$ versus $\bar{A}$. While there is no direct correlation between these two quantities, an interesting behavior does appear. There seems to be a 'jump' in the data such that liquidus temperatures below 600 K have values of $\bar{A} \leq 0.1$ and systems with liquidus temperatures greater than 800 K have values of $\bar{A}$ approximately $\geq 0.075$. The exact meaning of this behavior is unclear. The first real correlation appears when examining the relationship between $\bar{A}$ and $T_r$. It is evident from panel (b) of Fig. (\ref{Corr.}) that there exists a strong correlation between these quantities. Making this more quantitative, the value of the Pearson's correlation coefficient between $\bar{A}$ and $T_r$ was $r \approx 0.8$. This result is very interesting, and perhaps not surprising. 
\begin{figure*}
\centering
\includegraphics[width= 1.8 \columnwidth,keepaspectratio]{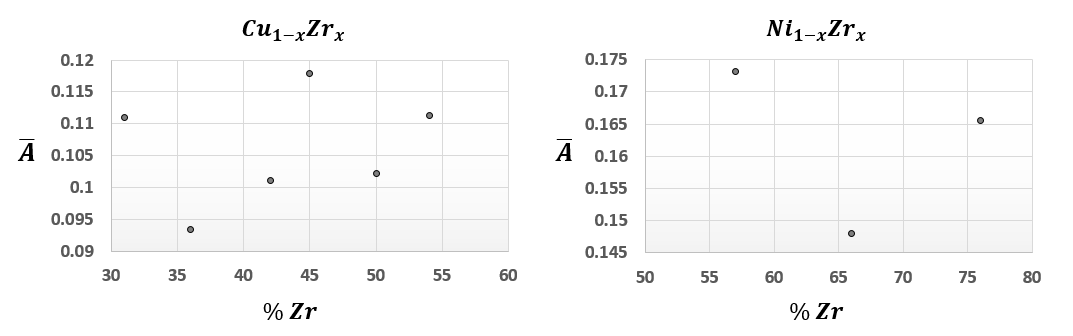}
\caption{(Color Online.) Correlation of $\bar{A}$ with the zirconium content of the binary systems $Cu_{1-x}Zr_x$ and $Ni_{1-x}Zr_x$. The CuZr system has well defined peaks in the value of $\bar{A}$ where it has been shown the expansivity also has peaks.}
\label{Zirc.}
\end{figure*}
In the years since Turnbull's original paper \cite{Turnbull}, the reduced glass transition temperature, $T_r$, has been used as a measure of glass forming ability in metallic liquids. The Turnbull temperature essentially quantifies how wide of a temperature range a liquid has to avoid crystallization.  It is clear from our data, that higher levels of  glass forming ability (GFA) correspond to smaller values of $\bar{A}$. This is consistent with the DEH framework. As the temperature is lowered, and the distribution over eigenstates shifts to lower energy states (the displacement of the distribution peak is governed by the heat capacity and is discussed in the next section). 

In addition to making good physical sense, and helping to facilitate an understanding of the physics underlying parameter, $\bar{A}$, this correlation may, in fact, allow for the prediction of the viscosity in the supercooled range. Using the equation of the linear fit in panel (b) of Eq. (\ref{Corr.}), $\bar{A}=0.268-0.2806T_r$. This enables a calculation of $\bar{A}$ from $T_g$ and $T_{l}$. This, in turn, enables the prediction of the temperature dependence of the viscosity in the undercooled regime. Thus, if $T_g$ can be measured in a thermodynamic sense, through calorimetric measurements of the heat capacity or specific volume, then the kinetic properties of the system can be entirely determined up to some error. Additionally, this would also potentially allow for the prediction of the liquidus temperature (associated with the dominant crystal phase), if one were to replace the parameter $\bar{A}$ in the fitting function with the reduced temperature, and fit with $T_{l}$ as the parameter. Both of these predictive abilities would represent major advances for the field of supercooled liquids.

For many of the metallic liquids studied, thermodynamic data related to the density and its temperature dependence were available at the liquidus temperature. We examined the correlation of $\bar{A}$ with the density at the liquidus, ${\it{p_{T=T_{l}}}}$, rate of change of density with temperature at the liquidus, $\frac{d{\it{p_{T}}}}{dT}|_{T_{l}}$, expansion coefficient at the liquidus, $\alpha(T_{l})$, and the scaled rate of change of density with temperature at the liquidus, $\frac{T_{l}}{{\it{p_{T=T_{l}}}}}\frac{d{\it{p_{T}}}}{dT}|_{T_{l}}$ for 14 metallic glass forming liquids. We saw no discernible correlation with either the density or expansion coefficient evaluated at the liquidus. Correlations with the rates of change of density are presented in panels (a) and (b) of Fig. (\ref{Corr2.}). From the figure, a quantifiable correlation between $\bar{A}$ and the rate of change of the number density, ${\it{p_{T}}}$ (both bare and scaled) at the melting temperature seems apparent. This result indicates that $\bar{A}$ may in fact be linked to equilibrium thermodynamic values. More strikingly, the rate of change of density at the melting point, is connected to both the nature of the potential for the given liquid, as well as the way the system `jams' as it is cooled. This not only suggests that $\bar{A}$ is linked to the microstructural interactions, but also that it may be able to connect the DEH with other concepts such as free volume, unifying many of the theories of the glass transition under the DEH `umbrella'. The DEH naturally rationalizes these, and many other experimentally observed trends. At a given temperature, $T$, a larger value of $\bar{A}$ means there is a wider distribution of states. If the equilibrium expansivity is temperature dependent (i.e. different for low and high energy eigenstates), then as temperature is varied, contributions from a broader range of these will lead to a greater rate of change of density, consistent with the observed results.

Finally, it was demonstrated in \cite{Expansivity}  that the expansivity for a range of compositions in the  $Cu_xZr_{1-x}$ system, possesses well defined local maxima at specific zirconium fractions. This intriguing result inspired us to analyze the behavior of $\bar{A}$ as a function of the zirconium fraction of supercooled metallic liquids in the $Cu_xZr_{1-x}$ and $Ni_xZr_{1-x}$ systems. These results are presented in Fig. (\ref{Zirc.}). From the figure, it is apparent, and surprising, that $\bar{A}$ also seems to 'oscillate' with well defined local maxima at various zirconium fractions. Preliminary results suggest these maxima may occur at similar zirconium fractions as the results in \cite{Expansivity} for the $Cu_xZr_{1-x}$ systems. The exact meaning of these results will be considered in a future work.

\subsection{Temperature Dependence of ${\it{p_{T}}}$: $\bar{A}$ and Heat Capacity}
\begin{figure}
\centering
\includegraphics[width=1 \columnwidth]{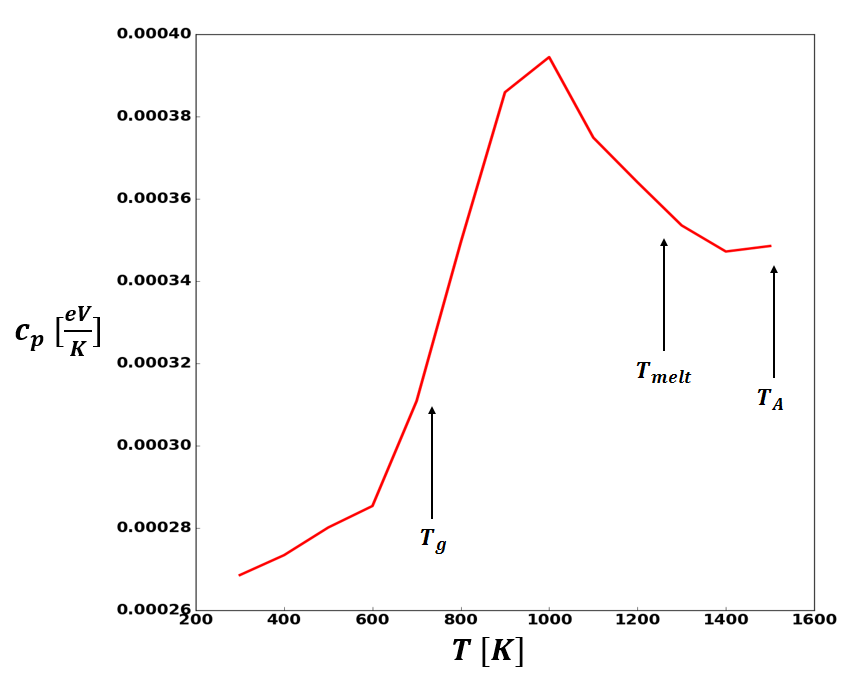}
\caption{(Color Online.) Heat capacity at constant pressure, $c_p$, for $Cu_{64}Zr_{36}$ system. (Data take from MD simulations \cite{Nick2}).}
\label{Heat.}
\end{figure}

In our discussion of the connection between the GFA and the DEH parameter, $\bar{A}$, we alluded to the link between the heat capacity and the thermal displacement of the distribution peak. The heat capacity, which is defined as $C_x\equiv\frac{dE}{dT}|_x$, controls the rate at which the average system energy (which is the same as the peak location for a Gaussian distribution) changes with temperature. It is important to point out that  $C_{e.q.}=\frac{dE}{dT}$ is the equilibrium heat capacity of the system and that in equilibrium the distribution is presumed $\delta$-peaked. Therefore, the equilibrium heat capacity quantifies the change in eigen-energy with decreasing/increasing temperature. The heat capacity of supercooled liquids, which are not in equilibrium, is the change in the \textit{average distributional energy} for distributions which are not $\delta$-functions. This average energy is the location of the peak of the distribution, and therefore the supercooled heat capacity, $C_{s.c.}=\frac{d<E>}{dT}$, controls the rate at which the center of the distribution over states changes with temperature, similar to how $\bar{A}$ determines the rate at which the width of the distribution changes. In general, supercooled liquid heat capacities exhibit a drastically different temperature dependence from that observed in the associated equilibrium system. As the liquid is supercooled (the temperature is lowered), the heat capacity shows an increase which begins at roughly, $T_A$, the dynamic crossover temperature. This increase continues monotonically with cooling up to a peak value at $T \approx 1.2 T_g$. Below this temperature, the supercooled liquid heat capacity drops down continuously to a value which approximates the value consistent with the vibrational heat capacity of the equilibrium crystalline solid for temperatures down to absolute zero. An example of this is shown in Fig. (\ref{Heat.}). 
\begin{figure*}
\centering
\includegraphics[width=1.8 \columnwidth]{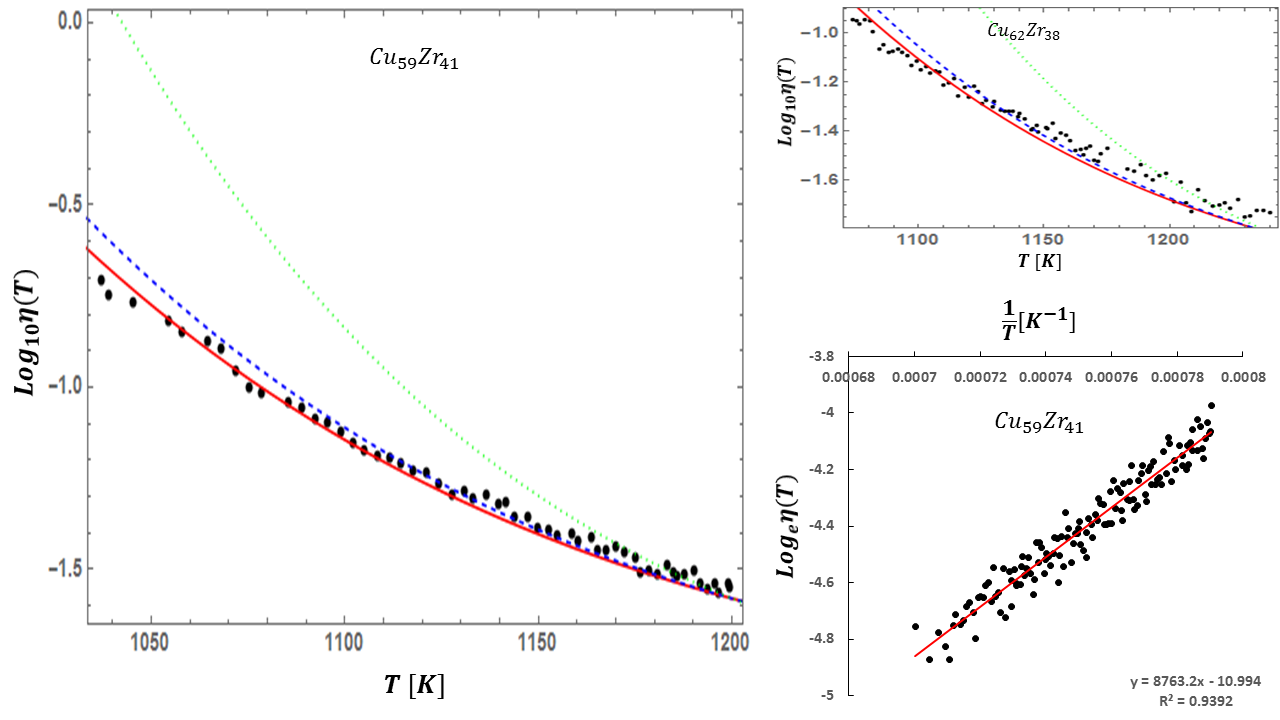}
\caption{(Color Online.) Viscosity data for two metallic liquids with the DEH viscosity function, Eq. (\ref{Final}) (solid red curve) applied. Using the correlation found in Figure (\ref{Corr2.}b) and the relationship between $E_{\infty}$ and $T_g$ discussed in Section VII, we predict the value of $\bar{A}$ from purely high temperature data as seen in the bottom right panel. The dashed, blue curve represents the fit using this predicted value. The green, dotted curve represents an `upper bound' on the prediction by considering the average error in the correlation involving $\bar{A}$ and high temperature measurements. Overall, the blue curves are seen to do a remarkable job representing the data at low temperatures using only high temperature measurements.}
\label{Est.}
\end{figure*}
We will now briefly discuss the implications of this temperature dependence of the heat capacity on the DEH model. The characteristic ``bump" which appears beneath $T_A$ means that for temperatures lower than $T_A$, the distribution begins to shift to lower energies increasingly rapidly. This effect combined with the decreasing width of the distribution implies that the Gaussian will begin to rapidly lose weight in the liquid-like states as the peak moves deeper into the regime of the solid-like states while the width simultaneously tightens. This will have the overall effect of rapidly decreasing the fluidity of the liquid, and provides a link to the non-Arrhenius character of the supercooled liquid viscosity. It has been observed that the size of the characteristic heat capacity bump in supercooled liquids varies in accordance with the spectrum of non-Arrhenius behaviors. This suggests that the combination of the size of the heat capacity bump and the parameter $\bar{A}$ will have some correlation to the fragility of the liquid. Taken together, these ideas suggest a strong link between thermodynamic quantities and kinetic quantities, a topic of intense debate in the glass community. In an upcoming paper, we will investigate this link fully, focusing on the thermodynamic and structural implications of the energy distribution framework, including a detailed microphysical picture that justifies the distribution, as well as links the behavior of the heat capacity to dynamic heterogeneities and structure.
\section{Prediction of $\eta(T)$ from High T Data}
In Section \ref{Corr}, it was demonstrated that $\bar{A}$ showed a strong correlation with the reduced glass transition temperature, $T_r$. It was suggested that knowing the value of $T_g$ and $T_{l}$ would then allow for the prediction of $\bar{A}$ and consequentially, the viscosity of the supercooled liquid. It may seem dubious that it would be necessary to use the glass transition temperature, which marks the lower limit of the supercooled regime, to predict the kinetics of the supercooled liquid. Additionally, it can be quite difficult to experimentally determine the value of $T_g$. By contrast, high temperature data, and associated features are often easier to measure in the laboratory. It would seem beneficial, then, to relate $\bar{A}$ to high temperature (melting and above) data. Recently,an empirical relationship between the glass transition temperature, $T_g$, and the high temperature activation energy, $E_{\infty}$, of the viscosity in the Arrhenius regime for metallic liquids was found \cite{Kelton2}. In that work, it was discovered that for all metallic liquids the relationship, $E_{\infty} \approx 11 k_B T_g$ holds. Taking advantage of this, as well as the equation of the linear fit, $\bar{A}=0.268-0.2806T_r$, resulting from the relationship displayed in panel (b) of Fig. (\ref{Corr.}), we can estimate the value of $\bar{A}$ for a metallic liquid within a bound associated with the average error in the fit. We applied this method to two metallic glass forming liquids, the results of which are depicted in Fig. (\ref{Est.}). In the figure, the red (solid) curve represents the fit to the data associated with the fitting as described in Section II. The blue (dashed) and green (dotted) curves represent the fits associated with the values of $\bar{A}$ on both sides of the predicted boundary. It is clear that the blue curve does a reasonable job of predicting the viscosity of the supercooled liquid using purely high temperature data. The specific values of $\bar{A}$ predicted will depend on how tightly the relationship of \cite{Kelton2} holds, as well as the sensitivity of the linear fit in the high temperature Arrhenius regime to the extraction of $E_{\infty}$. These results will need to be made more rigorous, but the preliminary results suggest that the viscosity of supercooled liquids can be predicted to reasonable accuracy using data which is more readily available than measuring the viscosity in the lab. These results will likely be of particular interest to researchers working in industry. To conclude this section, we point out that the relationship found in \cite{Kelton2} also holds for organic/molecular liquids, and a similar relationship for network formers, $E_{\infty} \approx m k_B T$, with m, the fragility, was suggested. This result was not able to be verified here, but represents an exciting research opportunity and possible extension of the DEH.

\section{The DEH Viscosity Above Melting}

Until now, we focused on the viscosity of supercooled liquids below their liquidus temperature, $T_{l}$. In Section \ref{the_others}, we contrasted the DEH with a series of alternative models for the viscosity of supercooled liquids, ultimately concluding that the DEH model more accurately described experimental data, with fewer parameters. These alternative models, however, apply to the entire temperature range in which viscosity measurements can be made. Therefore, it is imperative that the DEH model be able to make statements about the viscosity above melting. In deriving the DEH expression for the viscosity (Eq. (\ref{Final})), we began from the expression in Eq. (\ref{viscosity}). Using this expression and arguments laid out more concretely in \cite{Nussinov}, at temperatures above the liquidus, the viscosity is given by

\begin{eqnarray}
\label{highT}
\eta(T)=\frac{\tilde{\eta}}{\int_{T_{l}}^{\infty}~C_{v}^{'}(T^{'})r^{'}(T^{'}){\it{p_{T}}}(
E(T^{'}))~dT^{'}}.
\end{eqnarray}
Here, $\tilde{\eta}$ is a constant, $C_{v}^{'}$ is the constant volume heat capacity of the equilibrium liquid at a temperature $T^{'}$, while $r^{'} ( \propto \exp(-\frac{\Delta G(T^{'})}{k_B T^{'}}))$, and $E^{'}$ denote, respectively, the equilibrium relaxation rate and internal energy. As we have done earlier, we will, once again, invoke a Gaussian ${\it{p_{T}}}[E^{'}(T^{'})]\propto \exp(-\frac{(T^{'}-T)^2}{2 \bar{\sigma}_{T}^2})$. With these, Eq. (\ref{highT}) becomes
\begin{eqnarray}
\label{Full}
\centering
\eta(T)=\frac{\tilde{\eta}}{\int_{T_{l}}^{\infty}~C_{v}^{'}(T^{'}) e^{-\frac{\Delta G(T^{'})}{k_BT^{'}}}\frac{1}{\sqrt{2 \pi \bar{\sigma}_{T}^2}}e^{-\frac{(T^{'}-T)^2}{2 \bar{\sigma}_{T}^2}}~dT^{'}}.
\end{eqnarray}
For $T$ close to yet above the liquidus temperature, the distribution ${\it{p_{T}}}(
E
(T'))$ can still be assumed to be localized to a narrow range of $T'$ about the temperature, $T$. In this narrow range, $C_{v}^{'}(T')$ is essentially constant. As the system is supercooled, a temperature will eventually be reached where solid-like characteristics begin to set in. We have already met such a temperature, the crossover temperature, $T_A$. The integrand  vanishes approximately over a temperature range of order $\bar{\sigma}_T$ that is centered on $T \approx T_A$. When $\frac{\bar{\sigma}}{T} \ll 1$ (as is empirically the case), we may Taylor expand the heat capacity and free energy barrier about $T^{'}=T$, 
\begin{eqnarray}
\label{Taylor}
C^{'}_{v}(T^{'})\approx && C^{'}_{v}(T)+\frac{\partial C_{v}^{'}}{\partial T^{'}} (T^{'}-T)+ \cdots, \nonumber \\
\Delta G(T^{'}) \approx && \Delta G(T)+\frac{\partial \Delta G}{\partial T^{'}} (T^{'}-T) + \cdots \nonumber \\
= && \Delta H(T)-T\Delta S(T) \nonumber \\
&& +\frac{\partial \Delta H}{\partial T^{'}}(T^{'}-T)-T'\frac{\partial \Delta S}{\partial T}(T^{'}-T) + \cdots .
\end{eqnarray}
In Eq. (\ref{Taylor}), we used $\Delta G=\Delta H-T\Delta S$, where $H$ is the enthalpy (at a fixed volume, the enthalpy would be replaced
by the internal energy $E$), and $S$ is the entropy of the equilibrated system at a temperature $T'$. Noting that $\frac{\partial \Delta H}{\partial T'}=C^{'}_p(T')$ and $\frac{\partial \Delta S}{\partial T'}=C^{'}_v(T')/T'$, and using the fact that we are working in the regime where $T \approx T_A$, we have
\begin{eqnarray}
\label{Taylor2}
\Delta G(T') \approx && (\Delta G(T) - \Delta G(T_A)) +  \Delta H(T_A)-T\Delta S(T_A) \nonumber \\
&& +C^{'}_v(T_A)(T'-T)-T\frac{C^{'}_p(T_A)}{T}(T'-T) \nonumber
\\ &&+\cdots .
\end{eqnarray}
We recognize that $C^{'}_v(T_A)(T'-T)-T\frac{C^{'}_p(T_A)}{T}(T'-T)\approx (C^{'}_p(T_A)-C^{'}_v(T_A))(T'-T)=V(T_A)T_A\frac{\alpha(T_A)^2}{\beta(T_A)}(T'-T)\approx c_{1}(T'-T)$ (with $c_{1}$ a $T'$ independent constant).  Here, $\alpha(T_A)$ and $\beta(T_A)$ are, respectively, the thermal expansion coefficient and the isothermal compressibility at $T_A$. Setting $\Delta G(T_A)\equiv E_{\infty}$, the viscosity becomes
\begin{eqnarray}
\label{Full3}
\eta(T) \approx \frac{ \tilde{\eta}\sqrt{2 \pi\bar{\sigma}_{T}^2}  }{  C^{'}_v(T)  Ae^{-(\frac{\Delta G(T) - E_{\infty}}{k_{B} T_{A}})}
 \int_{T_{l}}^{\infty} e^{-c(T'-T) -\frac{E_{\infty}}{k_BT'} - \frac{(T'-T)^2}{2  \bar{\sigma}_{T}^2}} 
~dT'}. \nonumber \\
\end{eqnarray}
In Eq. (\ref{Full3}), $c= c_{1} + c_{2}$ where $c_{2} = (\Delta G(T) - E_{\infty})/(k_{B} T_{A}^{2})$.  
The Gaussian has its support in the narrow range of order ${\cal{O}}(\bar{\sigma}_T)$ about $T$, essentially forcing $T'$ to be close to $T$. 
If the three functions $\bar{\sigma}_{T}$, $C_v^{'}(T)$, $\Delta G(T)$ do not vary significantly in the interval $[T_{l},T_{A}]$, then for this range 
of temperatures, the viscosity
\begin{eqnarray}
\label{highTfinal}
\eta(T) \approx \eta_0 e^{\frac{E_{\infty}}{k_BT}}{e^{\frac{(T-T_A)^2}{2a^2}}}, 
\end{eqnarray}
where $a \equiv \bar{\sigma}_{T_A}$ and $\eta_0$ is a constant.
In Eq. (\ref{highTfinal}) we pulled out the Arrhenius factor of $e^{\frac{E_{\infty}}{k_BT}}$ (a factor that does not markedly change in the range $T_{A} > T > T_{l}$). This was done so that a trivial extension of Eq. (\ref{highTfinal}) that we write below will be valid at temperatures $T>T_{A}$. 
Far above $T_{l}$, the equilibrium Arrhenius form of Eq. (\ref{arrhenius}) follows from
Eq. (\ref{highT}) when $\Delta G(T')$ is weakly temperature dependent (and essentially equal to $E_{\infty}$). 
Thus, putting all of the pieces together, the viscosity for temperatures $T>T_{l}$ is, approximately,
\begin{eqnarray}
\label{HighFinal}
\eta(T) \approx \eta_0 e^{\frac{E_{\infty}}{k_BT}}{e^{\frac{(T-T_A)^2}{2a^2}\Theta(T_A-T)}} \nonumber \\
\equiv \eta_{equilibrium}(T){e^{\frac{(T-T_A)^2}{2a^2}\Theta(T_A-T)}}.
\end{eqnarray}
Similar to Eqs. (\ref{KKZNT}, \ref{BENK}), we explicitly inserted the Heaviside function $\Theta(x)$ that enables a crossover to the high temperature (Arrhenius) equilibrium form of the viscosity. We see, then, that the supercooled viscosity below $T_A$ is equal to the viscosity of the equilibrium liquid that is multiplied (and increased) by the reciprocal of a Gaussian This is reasonable since for temperatures below $T_A$, the weight $P_T(E')$ associated with the solid-like states (those states with energies $E'<E_{melt}$, the internal energy of the system at melting) will increase and the associated probability of a flow event will, correspondingly, decrease (and thus the viscosity will increase). 
\begin{figure*}
\centering
\includegraphics[width=1.8 \columnwidth]{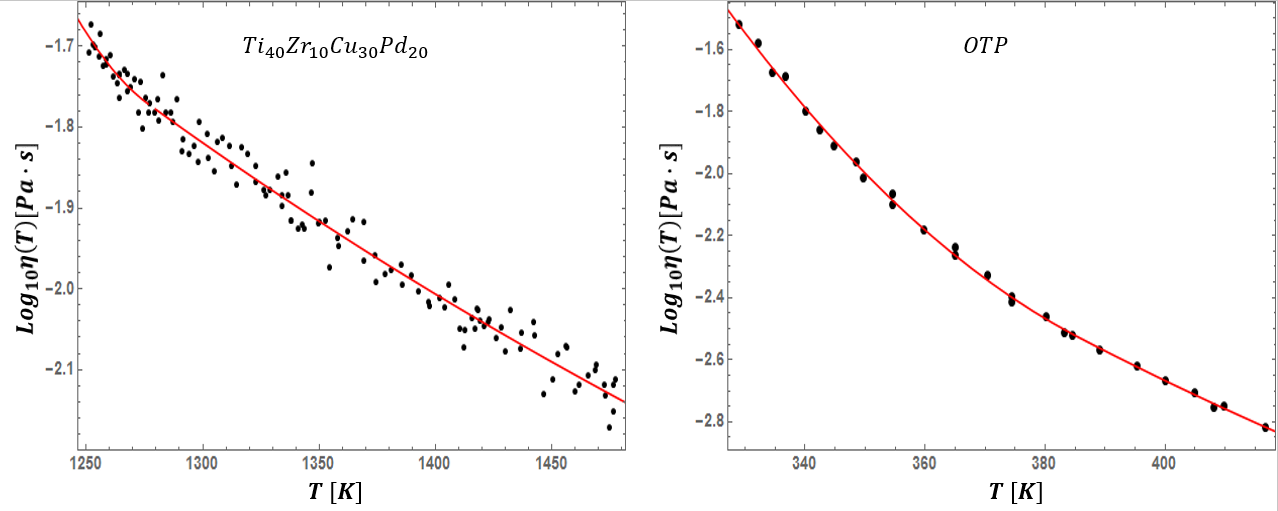}
\caption{(Color Online.) The high temperature form for the DEH viscosity, Eq. (\ref{HighFinal}) applied to experimental data of a metallic system (Ti$_{40 }$Zr$_{10}$Cu$_{30}$Pd$_{20}$) and the archetypal organic fragile glass former (o-Terphenyl (OTP)).}
\label{High.}
\end{figure*}
\begin{figure*}
	\centering
	\includegraphics[width=1.8 \columnwidth]{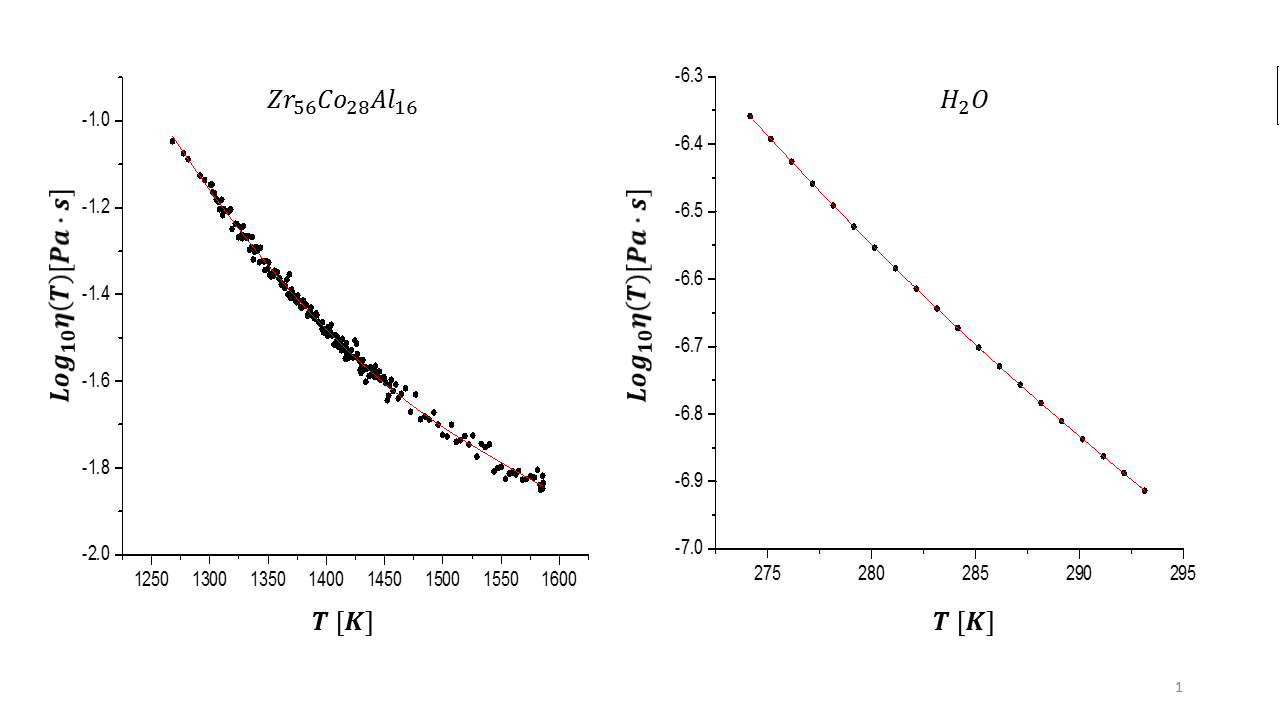}
	\caption{(Color Online.) The high temperature form for the DEH viscosity, Eq. (\ref{HighFinal}) applied to experimental data of a metallic system (Zr$_{56}$Co$_{28}$Al$_{16}$) and supercooled water.}
	\label{High2.}
\end{figure*}
It is worth pointing out that Eq. (\ref{HighFinal}) bears a striking resemblance to the BENK form \cite{Blodgett} of Eq. (\ref{BENK}) and the earlier associated parabolic fits of \cite{BENK1,BENK2}. Our derived viscosity of Eq. (\ref{HighFinal}) is exponential in a quadratic form in $T$ instead of $(1/T)$ as appears in \cite{BENK1,BENK2,Blodgett}. In these previous works, the parabolic functional form was presented without a theoretical framework justifying it. Here we directly derived Eq. (\ref{HighFinal}) within the DEH theory. Since the erfc appearing in Eq. (\ref{Final}) leads to an asymptotic Gaussian at low temperature \cite{Nussinov} and as at temperatures above the melting temperatures, we find the Gaussian form of  Eq. (\ref{HighFinal}), we  conclude that the pure Gaussian form appears in both the high and low temperature regimes of the DEH in a very physical manner. An added benefit of this form is that it holds down to $T_{l}$, and therefore, will allow for estimation of $\eta(T_{l})$ if such a value is not available. This makes using the form of Eq. (\ref{Final}) easier, without having to do an interpolation. 

Finally, we briefly demonstrate the quality of fit of Eq. (\ref{HighFinal}) to experimental data. Fig. (\ref{High.}) shows the viscosity data for both $Ti_{40}Zr_{10}Cu_{30}Pd_{20}$ and OTP with Eq. (\ref{HighFinal}) fit to the high temperature data. We leave all four parameters open, but note that we can easily constrain $E_{\infty}$ and $\eta_0$ with very high temperature data. This process can be volatile and prone to error with outliers from bad experimental data, so in this brief study we allow all parameters to be open. We see that the fit to $Ti_{40}Zr_{10}Cu_{30}Pd_{20}$ has $R^2=0.9776$ and $\chi_{reduced}^2=0.00219$ and OTP has $R^2=0.99879$ and $\chi_{reduced}^2=0.00111$, both indicating statsitically good fits. The predicted value of $T_A$ (1276-1286K and 386K, respectively) are reasonable, and the values for $a$, are such that $\frac{a}{T_A}<<1$, as assumed. 

\section{Jamming and other non-thermal transitions}
Glassy dynamics are ubiquitous in nature and appear in arenas that extend beyond the confines of the diverse collection of supercooled liquids  (wherein a rapid lowering of the temperature
drives the system into a glassy state) that we examined in the earlier sections of this paper. It is thus natural to investigate the possible links between the dynamics of non-thermal liquids and traditional supercooled liquids. As we will explain in this section, the formalisms of \cite{Nussinov,ESDH} are generally applicable to more than traditional supercooled liquids.
To make this lucid, we remark that the DEH eigenstates discussed hitherto (as well as the classical phase space regions of \cite{ESDH}) may not only be classified by the energy density or temperature but (in systems in which the volume and particle number are not both fixed) also by the number density and all other quantum numbers that describe them. The density matrix associated with the quenched, metastable system state will, generally, lead to an extension \cite{Nussinov} of the long time average of Eq. (\ref{average})
by an average with a probability distribution that depends 
on parameters other than the energy density (such as the volume fraction) if these parameters are allowed to vary. Similarly, in the classical approach 
of \cite{ESDH}, the long time average of Eq. (\ref{average}) will be performed over microstates that have different particle number and other parameters. 
Now, here is a new idea that we wish to introduce and explore in this section: if the quantum eigenstates or classical microstates change from being `liquid-like' to `solid-like'
as the number density (or other parameter) is increased then quenching will lead to a state for which much of our above ideas can be reproduced with 
the temperature $T$ replaced by the volume fraction (or other parameter) describing the macrostate of the system.

For concreteness, we will now explicitly consider the case of liquids which undergo a jamming transition \cite{He1,jam1,jam2,jam3,jam4,jam5,jam6,jam7,jam8,jam9,jam10,jam11,jam12,jam13,jam14}. The control parameter in this case is not the temperature but rather the volume fraction $\phi$. In, e.g., hard sphere (or colloidal) systems, $\phi$ is the fraction of the volume occupied by the hard spheres (or colloids). The system will again have a general Hamiltonian and associated eigenstates, and these eigenstates will intrinsically possess the macroscopic properties of equilibrium systems, including system sizes and atomic arrangement. For simplicity, the eigenstates can be associated with a many body energy, which will in turn depend on the volume fraction, $\phi$, instead of the temperature, $T$. Similar to the supercooled case, if the system starts out at as an equilibrated liquid (at volume fractions $\phi<\phi_{melt}$) and the volume fraction is then quasi-statically elevated then the system will remain in equilibrium and will transition to the crystalline state (when the volume fraction $\phi  = \phi_{melt}$). By contrast, if the system is very rapidly compressed, it will be driven out of equilibrium, and by the same arguments that we provided here and in \cite{Nussinov,ESDH} for supercooled liquids, the distribution ($p_{\phi}(E')$) will no longer be a $\delta$-function, but rather a (Gaussian) distribution parameterized by $\phi$ over the energy states that is of finite width. Thus, following rapid quenching, the quantum density matrix (or classical probability density) may generally contain both liquid-like and solid-like states. This mixed character will lead to the observed sluggish/glassy dynamics.
\begin{figure*}
	\centering
	\includegraphics[width=1.8 \columnwidth]{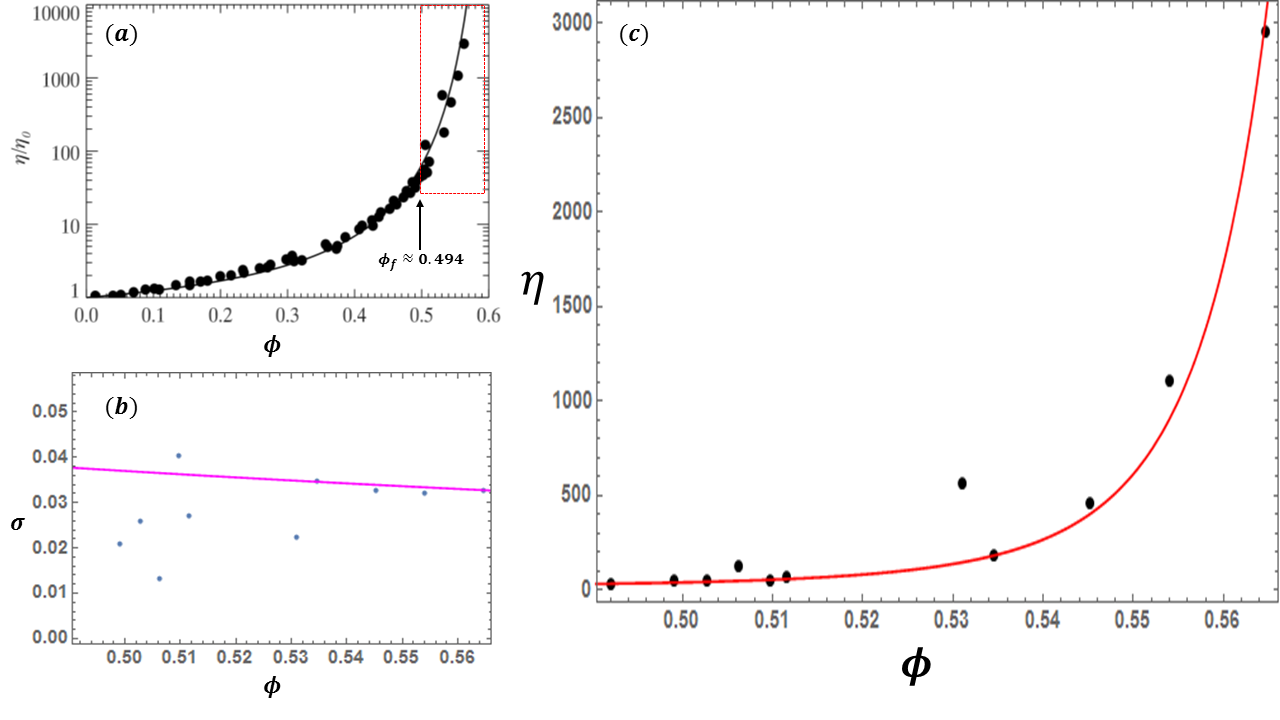}
	\caption{(Color Online.) (a) Reproduction of scaled experimental hard sphere viscosity data as a function of volume fraction, from \cite{JamData}. The equilibrium `freezing' volume fraction is marked and the metastable `supercooled/pre-jamming' region is highlighted in the dashed, red box. (b) Plot of the spread, $\sigma_{\phi}$, for the hard sphere data, found by inverting Eq.(\ref{jam1}). The magenta line is the curve $\sigma_{\phi}=\frac{\bar{J}}{\phi}$ using the value of $\bar{J}$ obtained from fitting Eq.(\ref{jamming}) to the data. (c) Hard sphere viscosity data with DEH fit applied. The DEH model is seen to do an exceptional job of reproducing the viscosity of a jammed hard sphere liquid, demonstrating the universality of the underlying physics of disordered solids.}
	\label{Jam.}
\end{figure*}
Assuming that only liquid-like states are capable of flow, such that the melting energy marks a cut-off and replicating, {\it mutatis mutandis}, the earlier steps that led to Eq. (\ref{viscosity}), we arrive at an functional form for the viscosity of jammed liquids as a function of the energy that is identical to that of the supercooled liquids that we discussed earlier, namely
\begin{eqnarray}
\label{jam1}
\eta(T) = \frac{\eta_{jam}(E_{melt})}{{\rm erfc} \Big( \frac{E_{melt}-\langle E \rangle}{\bar{\sigma}_{E} \sqrt{2}} \Big)}.
\end{eqnarray}
In order to make predictions, we need to convert this equation from being a function of energy levels to that of the relevant control parameter. In the case of jamming, the control parameter is the volume fraction, $\phi$ and not the temperature, $T$. We briefly sketch how, in the simplest approximation, the energy depends upon the volume fraction to make the necessary conversion. In the supercooled liquid case, the energy and temperature were simply related via an effective average heat capacity by $E=CT$. In systems where the volume can change, with fixed particle number, the energy changes correspond solely to volume changes (at constant pressure) and the energy $E=-PV$. The volume $V$ and the volume fraction $\phi$ are reciprocally related to one another ($V = \frac{const.}{\phi}$). Thus, in terms of volume fraction, the energy is $E=-\frac{const. \times P}{\phi} \equiv -\frac{\mathcal{P}}{\phi}$. Insertion of this relation into Eq. (\ref{jam1}) leads to
\begin{eqnarray}
\label{jam2}
\eta(\phi) = \frac{\eta_{jam}(\phi_{melt})}{{\rm erfc} \Big(( \frac{1}{\phi_{melt}}-\frac{1}{\phi})\frac{\mathcal{P}}{\bar{\sigma}_{E} \sqrt{2}} \Big)} \nonumber
\\
=\frac{\eta_{jam}(\phi_{melt})}{{\rm erfc} \Big( \frac{\mathcal{P}~(\phi-\phi_{melt})}{\bar{\sigma}_{E} \sqrt{2} \phi \phi_{melt}} \Big)}.
\end{eqnarray}
Since the energy $E$ and volume fraction $\phi$ are inversely related to one another, for small standard deviations (which we implicitly assume), we asymptotically have that $\sigma_E\approx\frac{\mathcal{P}\sigma_{\phi}}{\phi^2}$. If we further postulate that $\sigma_{\phi}=\frac{\bar{J}}{\phi}$ where $\bar{J}$ is a small material-dependent constant (similar to $\bar{A}$), then we finally obtain that 
for $\phi\geq\phi_{melt}$ the viscosity of the jammed fluid is
\begin{eqnarray}
\label{jamming}
\eta(\phi)=\frac{\eta(\phi_{melt})}{{\rm erfc}\left(\frac{(\phi - \phi_{melt})\phi^2}{\sqrt{2} \bar{J} \phi_{melt}}\right)}.
\end{eqnarray}
Similar to our earlier relation of Eq. (\ref{Final}), the viscosity of Eq. (\ref{jamming}) utilizes only a single parameter ($\bar{J}$), which controls the rate at which the width of the distribution changes with $\phi$, and a thermodynamically measured `freezing' point. 
In Fig. (\ref{Jam.}) we fit the viscosity function of Eq.(\ref{jamming}) to hard sphere data taken from \cite{JamData}. As seen in the figure, the DEH model is capable of very accurately reproducing the viscosity of the hard sphere liquid in the metastable, pre-jammed state. This result is highly significant, as it demonstrates the generality of the energy-distribution framework to different classes of amorphous solids, and provides a link between the jamming and glass transitions. Further, it begs the question as to whether phenomena such as shear thickening/thinning could also be explained using this framework with similar resulting functional dependencies of the viscosity on parameters such as applied stress. More investigation is required to answer this question, and it will be addressed in an upcoming paper. 

\section{Conclusion and Outlook.}
In this work, we expounded on a new framework for understanding supercooled liquids and the glass transition. Crucially, we tested the predictions of this theory 
(the distributed eigenstate hypothesis (DEH)) by analyzing the viscosity of {\it all} currently known supercooled liquid classes. We demonstrated, both qualitatively and quantitatively, that the DEH model can capture the temperature dependence of the viscosity of all of these liquid types to a statistically significant degree using only a \textbf{single} fitting parameter. We established that the viscosity of 45 disparate supercooled liquids below their melting temperature can be collapsed onto a universal curve over 16 decades by using the single parameter, $\bar{A}$. Coincident with the theoretical premise underlying the DEH theory, we unveiled
correlations between this single parameter and various properties of supercooled liquids and glasses. Our results further strongly hint that it may be possible to predict viscosity of supercooled liquids below their melting temperature using only viscosity data above this temperature. Notably, we also derived a new form for the viscosity above melting and assessed the validity of this functional form by examining experimental data. Taken together, our results suggest an underlying universality of the glass transition that enables a natural crossover from an activated Arrhenius form at high temperatures to a very marked rise of the viscosity (most pronounced in fragile systems) at low temperatures. While the predictions of the DEH framework led to our analysis and observations, it is possible that other approaches might also rationalize and complement our findings. We hope that our observations of {\it universal} behaviors in all known supercooled liquid types will spur further investigations.

\section{Acknowledgements} NW and ZN were supported by the NSF DMR-1411229.  ZN thanks the Feinberg foundation visiting faculty program at Weizmann Institute. CP and KFK were supported by the NSF DMR 12-06707, NSF DMR 15-06553, and NASA-NNX 10AU19G. NW would like to sincerely thank Robert Ashcraft and Rongrong Dai for supplying data and for very stimulating discussions. We would further like to thank Hajime Tanaka and Takeshi Egami for their kind words and encouragement.

\clearpage

\section{Supplementary Information}
\setcounter{figure}{0}
\setcounter{section}{0}
\makeatletter
\renewcommand{\thefigure}{S\arabic{figure}}
\renewcommand{\thesection}{S\Roman{section}}
\begin{figure*}
\centering
\includegraphics[width=1.8 \columnwidth]{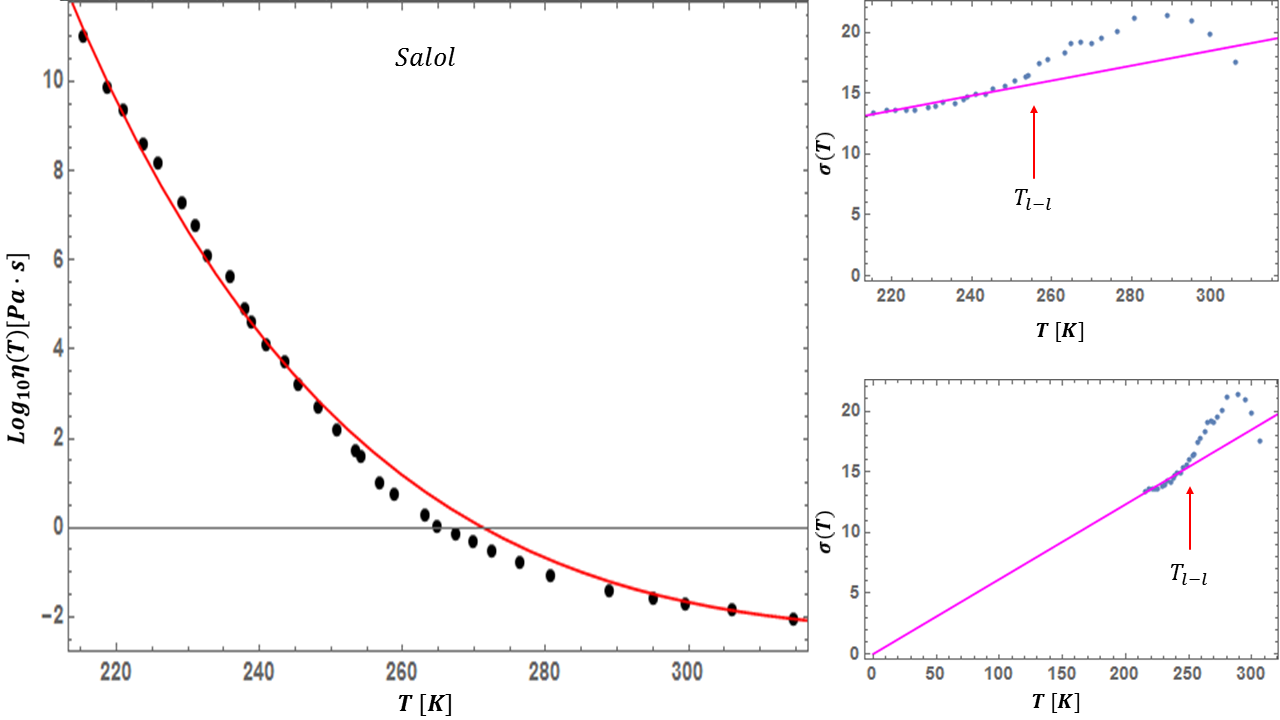}
\caption{(Color Online.) Left: DEH fit to the viscosity of Salol. The intermediate region of data where the DEH fit appears to "fail" corresponds to the region where the linear approximation breaks down and, in fact, to a temperature where a putative liquid-liquid phase transition was earlier suggested to occur (see text).
Right: (Top) The temperature dependence of $\bar{\sigma}_{T}$. The linear approximation fit from Eq. (\ref{viscosity}) (shown in magenta) works well over a large range of temperatures, but appears to break down upon approach to $T \approx 256K$ from below. (Bottom) $\bar{\sigma}_{T}$, this time with the range extended to the origin.}
\label{Salol.}
\end{figure*} 
\subsection{Exceptional Cases: Poor Fits and Possible Anomalies}
\label{sec:anomaly}

\begin{figure*}
\centering
\includegraphics[width= 1.8 \columnwidth, height= 0.5 \textheight, keepaspectratio]{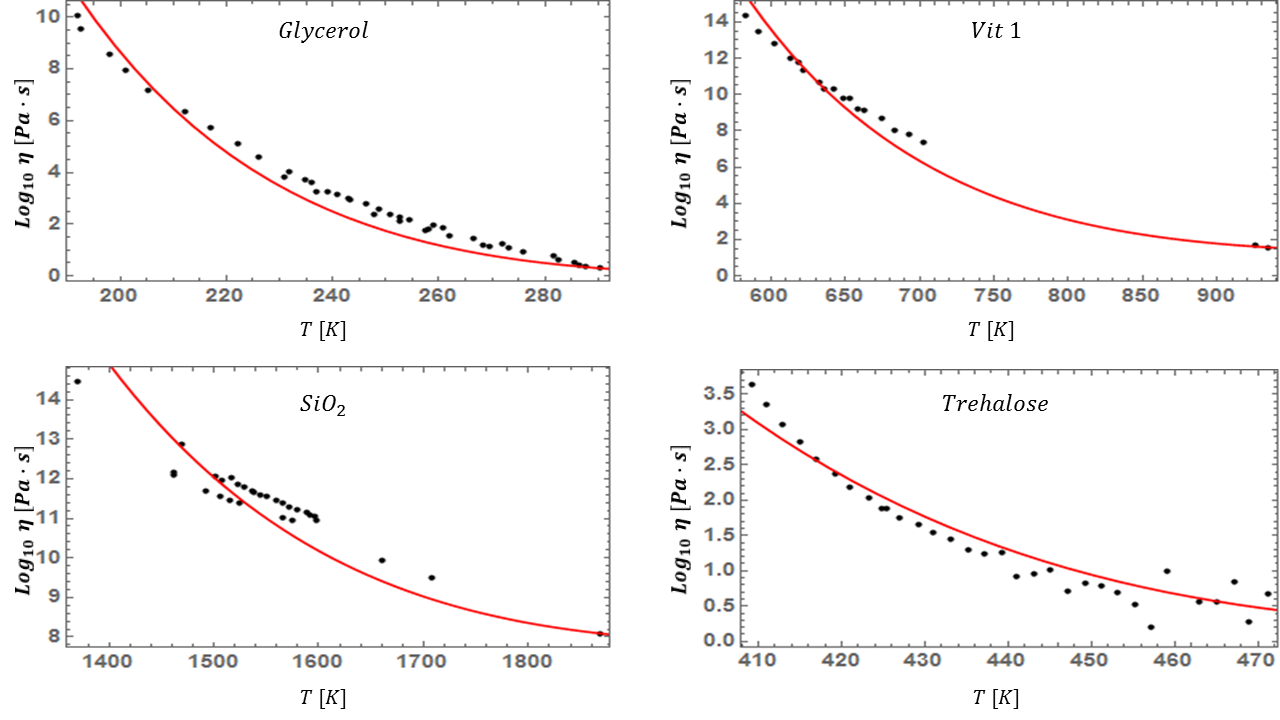}
\caption{(Color Online.) The four examples of worst performance of the DEH fit of Eq. (\ref{Final}). Various possibilities for the relatively poor performance are discussed above.}
\label{Bad.}
\end{figure*}

In the text it was demonstrated that the DEH form for the viscosity provided a statistically 'good' fit to the experimental data for a disparate group of supercooled liquids. Furthermore, we were able to collapse the data of all these liquids to a universal curve, representing both an underlying universality in the glass transition phenomenology, and a universality in the DEH formalism. It must be pointed out, however, that a few exceptional cases did appear in our analysis, with the DEH form not providing a reasonable fit to the data, despite all but one of the liquids being collapsed onto the universal curve. In Figs. (\ref{Salol.} and (\ref{Bad.}), data are presented for 5 exceptional case liquids with the DEH form applied, and it is clear that the function of Eq. (\ref{Final}) does not accurately describe the data as depicted. A multitude of reasons may exist for this discrepancy. In the cases of Vit 1 and Trehalose it seems likely that there is measurement error in the data. In the case of SiO$_2$, in addition to the pronounced scatter in the data, this liquid is also at the very extreme of strong behavior, and is traditionally difficult to describe with models. We are not sure exactly why this is the case. In the case of glycerol, there is data \cite{AngellGlyc} showing that the heat capacity has drastically different behavior from other supercooled liquids, and that a so-called liquid-liquid phase transition may exist in this liquid. Salol possesses similar anomalies. It was shown in \cite{Salol}, that Salol may undergo a so-called ``fragile-strong" crossover in the range $T_g \leq T \leq T_{l}$. This may be a liquid-liquid ($l-l$) phase transition, and the exact temperature of its occurrence was suggested to be at $T_{l-l}$=256 K \cite{Salol}. A system undergoing a phase transition or phase separation would not be expected to be able to be fit by a single form with a single parameter, which could lead to the discrepancy in the Salol and glycerol fits. The most interesting facet of this possible transition is that in examining the behavior of $\bar{\sigma}_T$ in Figure (\ref{Salol.}) for Salol, it is clear the the linear approximation breaks down at exactly the same temperature at which the putative liquid-liquid transition occurs, $T \approx 256 K$! Therefore, it may be possible for the DEH form to predict the existence and location of liquid-liquid phase transitions or crossovers, based on a change in the behavior of $\bar{\sigma}_{T}$. In fact, we observed that for the strongest liquids, $\bar{\sigma}_{T}$ tended to have a negative slope, instead of a positive one that the fragile liquids possessed. Therefore, crossing over from a negative slope to a positive one, or vice versa, may the signature of a strong-fragile transition. 

The reader will have noticed that the behavior of $\bar{\sigma}_{T}$ for glucose shows a stark crossover from increasing to decreasing at a sharp temperature in the supercooled range, yet the DEH model accurately fit the experimental data. It is possible that glucose may have a liquid-liquid transition, but this is masked in the DEH model, as the slope in the increasing range has the same magnitude in the decreasing range. The fact that the magnitude of the slope remained roughly the same will allow the DEH form to accurately predict the experimental viscosity data. 

Overall, many reasons may exist for the discrepancies observed in the fits of the above 5 liquids, ranging from experimental error to liquid-liquid phase transitions. The exact reasons will require further investigation, and will need to be understood to strengthen the validity of the DEH model.

\begin{thebibliography}{38}
\bibitem{qe} For completeness, we note that in recent years, a growing body of research has focused on apparent connections between traditionally quantum ideas/quantities and  observables (e.g., the viscosity) of high energy/temperature ``classical'' systems \cite{Bound}. In the current context, a link has been posited between the prefactor, $\eta_0$ in the high temperature Arrhenius expression of Eq. (\ref{arrhenius}) for the viscosity, and Planck's constant \cite{NussTome,Eyring,EyringBook,Blodgett}. Further suggestions have been raised of quantum mechanics playing a role in certain ``special liquids" \cite{Reichman}. Quantum jamming transitions \cite{qj} may might appear in systems in which the interatomic separations is of the order of the relevant thermal wavelength.
\end{thebibliography}
\end{document}